\documentclass[%
reprint,
superscriptaddress,
amsmath,amssymb,
aps,
]{revtex4-1}

\usepackage{graphicx}
\usepackage{dcolumn}
\usepackage{bm}
\usepackage{natbib}
\usepackage[colorlinks=true,citecolor=blue,linkcolor=blue,pdfhighlight =/O]{hyperref}
\hypersetup{urlcolor=blue}

\begin{document}


\title{ \Large Quantitative magneto-optical investigation of superconductor/ferromagnet hybrid structures}

\author{G. Shaw}
\altaffiliation{These authors contributed equally to this work.} 
\author{J. Brisbois} 
\altaffiliation{These authors contributed equally to this work.}
\affiliation{Experimental Physics of Nanostructured Materials, Q-MAT, CESAM, Universit\'{e} de Li\`{e}ge, B-4000 Sart Tilman, Belgium}

\author{L. B. G. L. Pinheiro}
\affiliation{Departamento de F\'{i}sica, Universidade Federal de S\~ao Carlos, 13565-905, S\~{a}o Carlos, SP, Brazil}
\affiliation{IFSP –- Instituto Federal de S\~ao Paulo, Campus S\~ao Carlos, 13.565-905, S\~ao Carlos, SP, Brazil.}

\author{J. M\"{u}ller} 
\author{S. Blanco Alvarez} 
\affiliation{Experimental Physics of Nanostructured Materials, Q-MAT, CESAM, Universit\'{e} de Li\`{e}ge, B-4000 Sart Tilman, Belgium}

\author{T. Devillers} 
\author{N. M. Dempsey} 
\affiliation{Univ. Grenoble Alpes, CNRS, Institut N\'{e}el, 38000, Grenoble, France}

\author{J. E. Scheerder} 
\author{J. Van de Vondel}
\affiliation{Laboratory of Solid--State Physics and Magnetism, KU Leuven, Celestijnenlaan 200 D, box 2414, B--3001 Leuven, Belgium}

\author{S. Melinte}
\affiliation{ Universit\'{e} catholique de Louvain, Institute of Information and Communication Technologies, Electronics and Applied Mathematics (ICTM), Institut de la Mati\`{e}re Condens\'{e}e et des Nanosciences (IMCN), 1348 Louvain-la-Neuve, Belgium}

\author{P. Vanderbemden}
\affiliation{ Universit\'{e} de Li\`{e}ge, SUPRATECS and Department of Electrical Engineering and Computer Science, Sart Tilman, B-4000, Belgium.}

\author{M. Motta}
\author{W. A. Ortiz}
\affiliation{Departamento de F\'{i}sica, Universidade Federal de S\~ao Carlos, 13565-905, S\~{a}o Carlos, SP, Brazil}

\author{K. Hasselbach} 
\author{R. B. G. Kramer} 
\affiliation{Univ. Grenoble Alpes, CNRS, Institut N\'{e}el, 38000, Grenoble, France}

\author{A. V. Silhanek} 
\affiliation{Experimental Physics of Nanostructured Materials, Q-MAT, CESAM, Universit\'{e} de Li\`{e}ge, B-4000 Sart Tilman, Belgium}

\date{\today} 
\begin{abstract} 
We present a detailed quantitative magneto-optical imaging study of several superconductor/ferromagnet hybrid structures, including Nb deposited on top of thermomagnetically patterned NdFeB, and permalloy/niobium with erasable and tailored magnetic landscapes imprinted in the permalloy layer. The magneto-optical imaging data is complemented with and compared to scanning Hall probe microscopy measurements. Comprehensive protocols have been developed for calibrating, testing, and converting Faraday rotation data to magnetic field maps. Applied to the acquired data, they reveal the comparatively weaker magnetic response of the superconductor from the background of larger fields and field gradients generated by the magnetic layer. 
\end{abstract}

\maketitle

\section{Introduction}
\label{sec_intro}
Magneto-optical imaging (MOI) has deservedly gained a privileged place among magnetic mapping techniques for investigating ferromagnetic\cite{McCord2015} and superconducting\cite{Jooss2002,Johansen2004} materials. The assets that make this technique stand out from the rest are its limited invasiveness\cite{Goa2003a,Brisbois2014,Brisbois2017}, short acquisition time\cite{Koblischka1995,Bending1999}, and a fair spatial and magnetic field resolution while being able to map the magnetic field profile over large surfaces. These advantages have allowed, for instance, to investigate the fast magnetic flux dynamics in superconductors\cite{Wertheimer1967,Bolz2003,Bujok1993} and large scale phenomena, such as flux avalanches\cite{Johansen2002,Brisbois2016}.

Over the years, a continuous strive to improve the MOI technique has allowed to reach a sufficiently high spatial resolution so as to achieve single flux quantum resolution\cite{Goa2001,Goa2003,Golubchik2009,Veshchunov2016}. Several protocols have been developed to enhance the magnetic field resolution of the MOI technique down to a few $\mu$T which has enabled the detection of subtle phenomena such as the melting of the Abrikosov vortex lattice\cite{Soibel2000,Banerjee2003,Wijngaarden2001,Mandal2012}. A major challenge in MOI remains the quantification of the acquired raw data\cite{Jooss2002}. Conversion of the light intensity distribution information obtained by MOI into magnetic field texture is complicated by several factors, such as non-uniform and time-dependent illumination, sample tilt and topography, depolarization by optical elements, defects and magnetic domains in the MOI indicator film, sample drift, intrinsic camera noise, etc. Hence, significant efforts have been made, first identifying and understanding the sources of limitations, and then mastering them for achieving a quantitative interpretation of MOI data\cite{Johansen1996,Laviano2003,Paturi2003,Roussel2007,Patterson2015,Albrecht2016,Grechishkin2016,Wells2016}.

Although MOI has been successful for mapping the magnetic field generated by magnets and superconductors, its performance in hybrid systems combining both type of materials is less certain. The reason being the difficulty to discern the contribution of each sub-system to the total magnetic field. In particular, when using hard ferromagnets and conventional superconductors, the former can lead to an overwhelming signal making it rather challenging to unveil the weak signal from the superconductor. One way to overcome this difficulty would be to quantify the field distribution in order to precisely remove the strong background signal from the ferromagnet. In this report, we present comprehensive protocols developed for conversion of the MOI data into a magnetic field distribution, with the goal of revealing the comparatively weaker magnetic response of a superconductor buried in a larger background field associated with a magnetic layer in its vicinity. Our work-flow, partially inspired by Ref. \onlinecite{Roussel2007}, involves calibration of intensities (gray levels) of images and conversion into local field values. For a superconducting sample, a set of calibration images are obtained above its critical temperature $T_\mathrm{c}$, providing magnetic field distribution information of the system in the absence of a superconducting signal. Then intensity versus field calibration curves are determined for each pixel in the image. The calibration curves thus obtained are subsequently applied on images acquired below $T_\mathrm{c}$ to obtain local magnetic field distribution information. The pixel-by-pixel calibration method enables an a posteriori correction of artifacts such as inhomogeneous illumination. Furthermore, the contrast originating from the in-plane magnetic domains in the indicator film has been removed by implementing a post image processing algorithm. The intensity-to-magnetic field conversion protocol and procedure for correction for magnetic domains are discussed in detail in Section \ref{sec_calib}, after introducing the employed experimental setup in Section \ref{sec_expt}. In the following Section \ref{sec_results} we discuss the application of the proposed protocols on systems of increasing complexity. In the first stage, we tested the protocols on MOI data obtained on magnetic structures (magnetic bars, disks, and a micro-scale electromagnet) with promising results (Sections \ref{sec_bars}, \ref{sec_dots}, and \ref{sec_coil}). Subsequently, we applied the technique to reveal new aspects of innovative superconductor/ferromagnet (S/F) hybrids (Sections \ref{sec_TMP} and \ref{sec_NbPy}). Quantification of MOI data is particularly useful for the hybrid structures as it enables clearer insight into the physics of these systems. However, as we discuss in Section \ref{sec_TMP}, we also confront the limitations of the proposed approach when applied on systems with large magnetic field amplitudes.

\section{Experimental details}
\label{sec_expt}

Let us start by introducing the experimental setup used for acquiring magneto-optical images. MOI is a magnetic field mapping microscopy technique based on the Faraday effect, where the direction of polarization of a light beam is rotated proportionally to the local magnetic field\cite{Jooss2002,Koblischka1995,McCord2015,Patterson2015,Lange2017}. This effect is strongest in purposely designed indicator films, placed on top of the sample under study. The Faraday active indicator we use throughout this work is a 3~$\mu$m thick Bi-doped yttrium iron garnet epitaxially grown on a 450~$\mu$m thick Gd$_3$Ga$_5$O$_{12}$ transparent substrate. A 100~nm thick Al mirror was deposited on the optically active layer side in order to assure sufficient reflection of the incident light beam produced by linearly polarizing the 550 nm emission of a Hg lamp. The magnetization of the Faraday active layer is in-plane in the absence of external magnetic field, but it is tilted out-of-plane by the presence of local magnetic fields perpendicular to the plane of the indicator. The rotation of the polarization is proportional to the component of the magnetic moment along the direction of light propagation. After crossing the analyzer, blocking the initial polarization direction, the reflected beam is captured by a high resolution RETIGA-4000 CCD camera recording 2048 px $\times$ 2048 px gray-scale images, thus obtaining a light intensity map representative of the magnetic field texture at the indicator's plane. In order to increase the signal-to-noise ratio, and unless stated otherwise, all the magneto-optical (MO) images presented throughout this work result from averaging between 3 and 10 images acquired by the camera with an exposure time of the order of 0.5 to 1 s. The polarization microscope is a commercial Olympus modular system, and the external magnetic field is provided by a cylindrical copper coil fed by a Keithley-2440 current source. Calibration of the coil was done with a USB Hall probe and consists in measuring the magnetic field as a function of the current at the center of the coil, at the location of the sample. Our configuration guarantees spatial variations of the field at most 1\% of the maximum field at the sample location (in a $5 \times 5$ mm$^2$ area). The sample is cooled down to temperatures as low as 4 K in a closed-cycle He cryostat (Montana Cryostation). The whole setup is installed on an actively damped non-magnetic Newport optical table. More details about the setup can be found in the supplementary material and in Ref. \onlinecite{Brisbois2016a}.

In some cases, MOI data has been complemented and compared with scanning Hall probe microscopy (SHPM) at room temperature. This technique allows acquisition of the real magnetic field $B_{z}(x,y)$ at various heights and was used to obtain direct quantitative information about the $z$-component of the stray magnetic field in the studied samples, over areas of hundreds of $\mu$m$^2$. A Hall probe with 5 $\times$ 5 $\mu$m$^2$ active area was used for the measurements. The scan resolution is 2.5 $\mu$m. Further details on the SHPM setup can be found in Ref. \onlinecite{Shaw2016}. 



\section{Quantitative magneto-optical imaging}
\label{sec_calib}

\subsection{Intensity-to-magnetic field conversion protocol}
\begin{figure*}[ht]
	\centering
	\includegraphics[width=16.5cm]{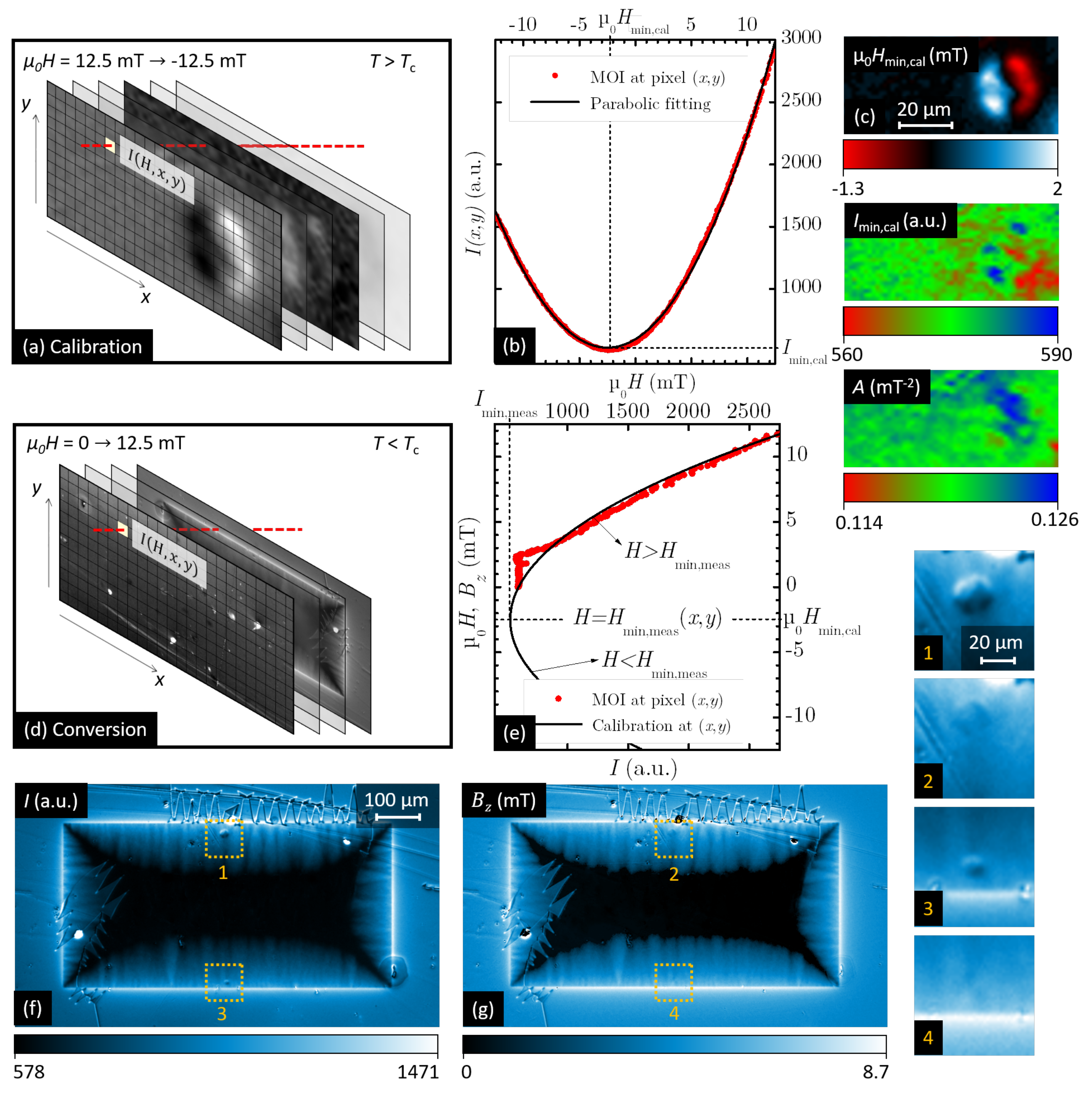}
	\caption{ {\bf Quantitative magneto-optical imaging.} (a) Data cube formed by the reference intensity images $I(x,y)$ recorded above $T_\mathrm{c}$ when sweeping the applied magnetic field $H$. (b) The intensity profiles $I(H)$ for each $(x,y)$ pixel are fitted by a parabolic curve, given by Eq. (\ref{eq:calib}), and have a minimum $I_\mathrm{min,cal}$ for an applied field $H_\mathrm{min,cal}$. These two parameters, as well as the concavity $A$, are mapped in panel (c) for a small part of the sample, including a $20 \, \mu$m diameter in-plane polarized magnetic disk. (d) Sequence of images to be converted to local magnetic field $B_{z}$, taken below $T_\mathrm{c}$ when sweeping $H$. (e) The $I(H)$ curve for each $(x,y)$ pixel is compared to the calibration curve to determine $B_{z}(I)$. To discriminate between the two possible values of $B_{z}$ for a given $I$, the field $H_\mathrm{min,meas}$ at which the minimum of intensity $I_\mathrm{min,meas}$ occurs is compared with the applied field $H$: if $H>H_\mathrm{min,meas}$ ($H<H_\mathrm{min,meas}$), the upper (lower) branch of the reference parabola is selected. A comparison of the original $I$ image with the final $B_{z}$ image is shown in panels (f-g) for a superconducting film with magnetic disks, at $T=3.75$ K and $\mu_0 H = 5.5$ mT. The enlargements on the right show the effectiveness of the conversion procedure to remove the inhomogeneous field source of the magnetic disks.}
	\label{Fig-calib}
\end{figure*}

MOI does not provide direct access to the out-of-plane magnetic field $B_z$, but rather to light intensity values $I$ related to it. Therefore, caution must be exercised when interpreting the raw images, since $I$ at a given pixel depends strongly on several parameters other than the local $B_z$, such as incident light intensity, exposure time, depolarization effects due to the optics, indicator tilt, or in-plane magnetic field components, to name a few. Furthermore, depending on the choice of the analyzer angle, pixels with the same light intensity may sometimes correspond to different values of the magnetic field. When it is desirable to recover information on the local magnetic field, protocols have been developed to convert the light intensity pixel values into absolute magnetic field values\cite{Jooss2002, Habermeier1979,Johansen1996,Laviano2003,Roussel2007}. The procedure we present here is inspired by those, but with two significant added values: (i) we use an exact pixel-by-pixel method\cite{Rave1987,McCord1999,Patterson2015}, instead of generalizing a calibration performed on a reference zone to the whole image, and (ii) we separate the contribution of the superconductor from other sources of constant magnetic field, such as ferromagnetic structures. The complete procedure is summarized in Fig. \ref{Fig-calib} using as illustration a rectangular superconducting Nb film with two ferromagnetic disks as sources of inhomogeneous magnetic field.

According to Malus' law, the intensity $I(B_z,x,y)$ recorded by the camera for a pixel $(x,y)$ in the image can be approximated as follows\cite{Jooss2002}:
\begin{equation}\label{eq:malus}
I(B_z,x,y) = I_0(x,y) \sin^2 \left( \alpha(B_z,x,y) + \beta (x,y) \right),
\end{equation}
where $I_0 (x,y)$ is the incident light intensity, diminished by depolarizing effects and absorption, $\alpha(B_z,x,y)$ is the angle of rotation of the light polarization, coming from the Faraday effect for a local out-of-plane magnetic field $B_z(x,y)$, and $\beta (x,y)$ is the deviation from the extinction configuration. The spatial distribution of $\beta (x,y)$ is non-uniform, mainly due to the small deviation of the incident light from normal incidence. This means that the extinction angle is not uniquely defined for a given image. In other words, the minimum of intensity does not occur for the same analyzer-polarizer angle at every pixel $(x,y)$. The value of $\alpha(B_z,x,y)$ is related to $B_z(x,y)$ through the out-of-plane magnetization $M_z(x,y)$ of the MO active layer of the indicator: $\alpha(B_z,x,y) = C(x,y) M_z(B_z,x,y)$, where $C(x,y)$ depends on the sensitivity of the MO active layer. Indeed, $B_z$ affects the originally in-plane magnetization of the MO active layer of the indicator by tilting it out of the plane. The out-of-plane component of the magnetization is given by $M_z(B_z,x,y) = M_{\mathrm{s}} \sin \theta(B_z,x,y)$, with $M_{\mathrm{s}}$ the saturation magnetization of the MO active layer and $\theta = \arctan \left( B_z(x,y)/B_{\mathrm{s}} \right)$ the angle between $M_\mathrm{s}$ and the in-plane direction\cite{theta}. $B_{\mathrm{s}}$ is related to the saturation field of the MO active layer and is of the order of 80 mT for indicators of the type employed in this work\cite{Johansen1996}. Substitution of $\alpha(B_z,x,y)$ in Eq. (\ref{eq:malus}) yields the following expression:
\begin{widetext}
\begin{equation}\label{eq:calib_exact}
I(B_z,x,y) = I_0(x,y) \sin^2 \left( C(x,y) M_{\mathrm{s}} \sin \left( \arctan \left( \frac{B_z(x,y)}{B_{\mathrm{s}}}\right) \right) + \beta(x,y) \right).
\end{equation}
\end{widetext}

This equation, linking $B_z(x,y)$ with the intensity $I(x,y)$ picked up by MOI, is at the core of quantitative MOI. Provided the parameters are determined through a preliminary calibration done on a set of data where $B_z(x,y)$ is known, Eq. (\ref{eq:calib_exact}) can be used to convert $I(x,y)$ images to $B_z(x,y)$ field maps in any subsequent measurements.

When local magnetic fields $B_z(x,y) \ll B_{\mathrm{s}}$, which is the case in most of our experiments (with the notable exception of the patterns presented in section \ref{sec_TMP}), one can make a Taylor series expansion of Eq. (\ref{eq:calib_exact}), keeping only terms up to order 2. In this case, $I(B_z,x,y)$ takes a much simpler parabolic dependence on $B_z(x,y)$:
\begin{equation}
I(B_z,x,y) \simeq a(x,y) + b(x,y) B_z + c(x,y) B_z^2,
\end{equation}
where 
\begin{align}
a(x,y) &=I_0 \sin (C M_{\mathrm{s}} \sin \beta),\\
b(x,y) &=\frac{I_0 C M_{\mathrm{s}}}{B_{\mathrm{s}}} \cos \beta \sin (2 C M_{\mathrm{s}} \sin \beta),\\
c(x,y) &=\frac{I_0 C M_{\mathrm{s}}}{B^2_{\mathrm{s}}} \lbrace C M_{\mathrm{s}} \cos^2 \beta \cos (2 C M_{\mathrm{s}} \sin \beta) \notag \\
 &- \frac{\sin \beta}{2} \sin (2 C M_{\mathrm{s}} \sin \beta) \rbrace.
\end{align}
This equation can also be rewritten in another form, so as the parameters acquire a more intuitive physical interpretation:
\begin{equation}\label{eq:calib}
I(B_z,x,y) \simeq I_\mathrm{min}(x,y) + A(x,y) \left[ B_z(x,y) - B_\mathrm{min}(x,y) \right]^2
\end{equation}
where $A(x,y)=c(x,y)$, $B_\mathrm{min}(x,y)=-b(x,y)/2c(x,y)$, and $I_\mathrm{min}(x,y)=a(x,y)-b^2(x,y)/4c(x,y)$, are experimental parameters to be determined for each $(x,y)$ pixel of the image. Indeed, in Eq. (\ref{eq:calib}), $B_\mathrm{min}(x,y)$ is the value of the local magnetic field that gives the minimum intensity $I_\mathrm{min}(x,y)$ for a given pixel $(x,y)$, while $A(x,y)$ is related to the sensitivity of the MO active layer.

The aim of the calibration procedure is to extract values of these parameters for each system configuration, i.e., for a given set of sample, indicator mounting, camera parameter and analyzer angle. This is done by recording a series of images, varying the external applied magnetic field $H=B_{z}/\mu_0$, as illustrated in Fig. \ref{Fig-calib}(a). In the case of a superconducting sample, the calibration needs to be done at a temperature $T>T_\mathrm{c}$, in order to exclude any superconducting signal. Under this condition, we typically record the average of 3 images for every value of $\mu_0 H$, varying between +12.5 mT and -12.5 mT by steps of 0.1 mT. Since we perform pixel-by-pixel calibration, it is important that all the recorded images overlap perfectly, i.e. unwanted drifts occurring during the measurements have to be taken into account. To that end, we use the StackReg plugin\cite{StackReg} of the ImageJ software\cite{ImageJ} which corrects the small translation of the sample of the order of $0.15 \, \mu$m/min (about 0.1 px/min), due to temperature gradients in the cryostat. The correction provided by the software has an error of $\pm 2$ px (i.e., about $3 \, \mu$m).

The sequence of calibration images thus obtained can be regarded as a data cube made of $N$ images with dimensions of $N_x \times N_y$ px$^2$, where each pixel contains the intensity value $I(H,x,y)$. Note that since the unwanted fluctuations of the light intensity produced by the Hg-lamp are homogeneous, they can be accounted for and eliminated by fitting with Eq. (\ref{eq:calib}) the intensity $I_\mathrm{mean}(H)$, corresponding to the average intensity in a $30 \times 30$ px$^2$ region free of defects and located far from the sample (i.e., where the magnetic field is homogeneous and has practically the same intensity as the applied field). Subsequently, the fitting of $I_\mathrm{mean}(H)$ returns the parabolic function $I_\mathrm{ref}(H)$, and for each image $I(H,x,y)$ is then corrected by multiplying every pixel by $I_\mathrm{ref}(H)/I_\mathrm{mean}(H)$.

After correcting for image drift and fluctuations of intensity, we plot the intensity $I(H,x,y)$ as a function of the applied magnetic field $H$ for each pixel $(x,y)$ of the data cube, and fit it with Eq. (\ref{eq:calib}), as illustrated in Fig. \ref{Fig-calib}(b). The calibration process yields the functions $H_\mathrm{min,cal}(x,y)$, $I_\mathrm{min,cal}(x,y)$ and $A(x,y)$, represented in Fig. \ref{Fig-calib}(c). This procedure allows us to accurately relate the detected intensity $I(H,x,y)$ with the local magnetic field $B_{z}(x,y)=\mu_0 H(x,y)$, using for each pixel its specific reference curve $I(H,x,y)$. This means that the calibration takes into account and allows to eliminate any spatial inhomogeneity in the illumination, defects and artifacts in the sample and indicator, as well as constant magnetic field sources other than $H$. Of course, we cannot recover magnetic field information from unresponsive pixels (for instance those resulting from damage in the indicator due to scratches) or when the light intensity saturates (due to local high magnetic fields), so the parameter values at these points do not have a physical interpretation. Interpolation procedures could be used as a first approach to estimate this missing information.

$H_\mathrm{min,cal}(x,y)$ represents the magnetic field at which the minimum of intensity $I_\mathrm{min,cal}(x,y)$ occurs. In the absence of fields other than the external applied field $H$, $H_\mathrm{min,cal}$ basically corresponds to the field rotating the light polarization by an angle $-\beta$, compensating the deviation of the analyzer and polarizer from the crossed configuration. However, in the presence of a magnetic field source other than $H$, the additional contribution to the rotation of light polarization also has to be compensated, thus impacting the value of $H_\mathrm{min,cal}$. Therefore, $H_\mathrm{min,cal}$ already provides us with a cartography of the magnetic field source, as can be seen in the example of the magnetic disk presented in Fig. \ref{Fig-calib}(c). Provided we subtract the background coming from $\beta(x,y)$, we have access to the average field generated by the source over the thickness of the MO active layer. Moreover, although the $I_\mathrm{min,cal}(x,y)$ distribution shows primarily the inhomogeneities in the background intensity, it may also reflect the shape of an inhomogeneous magnetic field source. This effect results from the fact that a uniform external magnetic field $H$ can not exactly compensate the local magnetic field $B$ generated by the inhomogeneous magnetic field source, since $B$ will be non-uniform through the thickness of the MO active layer, due to field decay with  distance. The partial compensation thus leads to a change in $I_\mathrm{min,cal}$. Finally, the parameter $A(x,y)$ is related to the sensitivity of the indicator, and may also vary due to composition inhomogeneities in the indicator. A careful look at Fig. \ref{Fig-calib}(c) allows us to identify some traces of the magnetic disk on the $A(x,y)$ parameter, which fall within the error bar of the procedure.

Let us now illustrate this calibration procedure in order to convert the collected light intensity distribution into magnetic field maps in a set of images corresponding to the rectangular Nb superconducting sample with two magnetic (Co) disks on top. More details of this particular sample layout are presented in section \ref{sec_dots}. At the end of the conversion procedure, the inhomogeneous field produced by the magnetic disks will be removed from the field map, leaving only the signal of the superconductor. For that purpose, we usually take a set of 10 images/field at $T<T_\mathrm{c}$ applying a zero-field-cooling (ZFC) procedure and increasing $\mu_0 H$ from 0 to 12.5 mT, as illustrated on Fig. \ref{Fig-calib}(d). As explained above, we correct the sample drift and the fluctuations of the incident light in the images, taking care of choosing exactly the same reference $30 \times 30$ px$^2$ region as for the calibration. Ideally, the $I(H)$ parabola in the reference square should be, after correction, the same as for the calibration set.

Subsequently, $I(H,x,y)$ curves for every pixel of the data cube are acquired (red curve in Fig. \ref{Fig-calib}(e)) and the value of the applied field $H_\mathrm{min,meas}$ at which the minimum intensity $I_\mathrm{min,meas}$ appears is estimated. This is an important step in order to convert the intensity into local magnetic field, since for each value of $I(H,x,y)$, there are two possible corresponding values of $B_{z}(x,y)$, one on each branch of the calibration parabola (black curve in Fig. \ref{Fig-calib}(e)). The criteria determining the proper $B_{z}$ for a given $I(H,x,y)$ is to compare $H$ with $H_\mathrm{min,meas}(x,y)$: if $H<H_\mathrm{min,meas}(x,y)$, the lower branch of the $B_{z}(I)$ curve needs to be used; if $H>H_\mathrm{min,meas}(x,y)$, the upper branch must be considered. For this reason, and to avoid imprecision on intensities close to the parabola minimum, it is convenient to choose in advance the analyzer angle appropriately $(\sim 4^{\circ})$ , if possible, so that the minimum of intensity does not occur in the range of applied fields $H$.

The result of the conversion procedure is shown in Fig. \ref{Fig-calib}(f-g), for the rectangular superconductor with magnetic disks highlighted by the yellow dotted rectangular selections. Figure \ref{Fig-calib}(f) shows the original light intensity image compared with the final quantitative magnetic field image in Fig. \ref{Fig-calib}(g). Note that throughout this report, the color scales in images are adjusted such that \textquoteleft up\textquoteright\space (\textquoteleft down\textquoteright) magnetization or \textquoteleft positive\textquoteright\space (\textquoteleft negative\textquoteright) magnetic field is represented by blue-white (red). Zero field is represented by black. After conversion, the background of the image is more homogeneous, and some inhomogeneities in the indicator response, such as the circular spot in the upper right corner of the sample, are corrected within the error associated to the method. Moreover, using this procedure, we can significantly attenuate the signal coming from ferromagnetic materials (with the assumption the signal does not change in the range of fields we apply) and separate it from the contribution due to the superconductor. In Fig. \ref{Fig-calib}, the magnetic disks are notably visible in panel (f), but become hardly visible in panel (g). This is confirmed by the direct comparison of the zoom-in of the disks in panel (f) (images 1 and 2) with the same regions in panel (g) (images 3 and 4). Unfortunately, magnetic domains and artifacts arising from the MO active layer change as the magnetic field is ramped, and thus can not be fully accounted for using this procedure. However, as we explain in the remainder of the section, it is still possible to remove them from the images to some extent. 

Note that our quantitative procedure does not take into account possible in-plane components of magnetic field that might affect the Faraday rotation in the indicator film\cite{Johansen1996,Laviano2003}. In-plane components are primarily induced by non-homogeneous magnetic field sources such as superconductors or ferromagnets. For the systems we have studied in the present work, we expect these components to lie below a few mT and therefore to be substantially smaller than the saturation magnetization of the indicator films ($\sim$ 80 mT). This justifies neglecting this effect and the associated corrections. Furthermore, consideration and estimation of in-plane components involves elaborate protocols of inversion of magnetic field maps to get current density distribution maps, which are then used to correct the magnetic field information in an iterative manner. The first iteration gives the out-of-plane component ($B_{z}$) only, and even after correction due to in-plane components, most of the features in $B_{z}$ are not altered significantly. As a consequence, the features we are interested in this report would not change noticeably by inclusion of these corrections.

\subsection{Correction for magnetic domains in the MO indicator}
\label{subsec_correction}

\begin{figure}[ht]
	\centering
	\includegraphics[width=6cm]{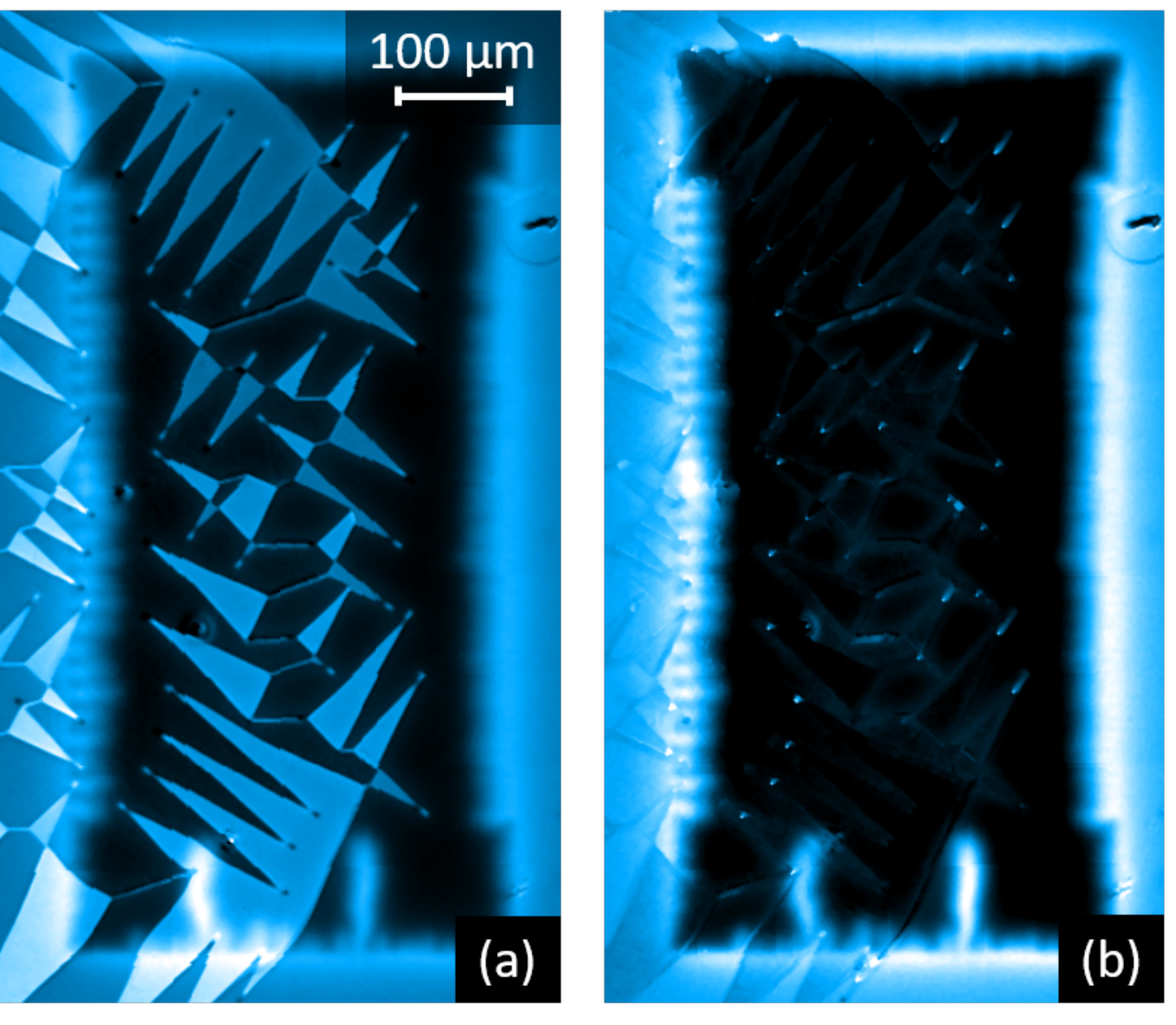}
	\caption{ {\bf Illustration of the correction of the indicator magnetic domains in a MO image.} (a) Original raw image of a $800 \times 400 \, \mu$m$^2$ Nb superconducting film at $T=3.6$ K and $\mu_0 H=2.1$ mT\cite{Brisbois2016a}, where the triangular artifacts pollute the intensity map. (b) Image after treatment with the correction algorithm, showing an enhanced contrast of the intensity map.}
	\label{Fig-corr}
\end{figure}

Special care has to be taken in order to avoid the proliferation of magnetic domain walls in the MO active layer of the indicator, since in their presence the local magnetic field is substantially modified and the technique can therefore no longer be considered as non-invasive\cite{Vestgaarden2007}. These magnetic domains appear as triangular-shaped regions where the intensity undergoes a jump compared to the neighboring regions thus degrading the clarity of the images and sometimes hiding actual sample features. In this section, we present an original method to reduce the impact of the magnetic domains on the images, by correcting the concerned areas. As a proof of concept, we apply our procedure in a situation where domain proliferation is favorable, which is the case when large gradients of magnetic field are present. This is clearly visible in the image shown in Fig. \ref{Fig-corr}(a), representing the magnetic field distribution in a rectangular superconducting Nb sample at $T=3.6$ K for a perpendicular applied field of $\mu_0 H=2.1$ mT.

The correction algorithm takes as input the 12-bit gray-scale image obtained from the experiment. First, the boundaries of the magnetic domains are drawn by applying a discrete differential operator (i.e., Sobel filter), based on the value of the intensity gradient at each pixel, as an edge detector. This operation creates a black and white boundary map of the image. In this map, we then select manually the regions corresponding to artifacts due to the indicator magnetization domains. We assume that the effect of a domain on the underlying image is to shift the intensity by a constant value, possibly different for every artifact\cite{shift}. In order to determine these constants, the user has to provide two reference points for each artifact, one inside and one outside the domain. The difference between the intensity at these two points gives the correction to apply.

Finally, using the boundary map drawn with the edge detector, we apply the appropriate correction to each domain using a flood fill algorithm. Since the corrections are based on a map of the boundaries and applied only to non-boundary pixels, this method leaves at least a single uncorrected pixel line between each domain. To smooth the image, an optional step is to replace each pixel of the boundary by the mean value of the nearest pixels not belonging to any boundary. This is however not always helpful as this step requires a precise determination of the boundaries, which cannot always be achieved, for instance in noisy images or for magnetically textured samples, where it could result in a degraded image quality. 

The result of the correction procedure described above is illustrated in Fig. \ref{Fig-corr}(b). As a final remark, since this method manipulates the values of the intensity, caution must be exercised when trying to extract quantitative information from the corrected image.

\section{Results and discussion}
\label{sec_results}

In the remainder of the paper, we apply the conversion protocol detailed above to study various ferromagnet/superconductor hybrid structures with increasing complexity. We start by analyzing a simple non-superconducting system made of magnetic Co bars and use it to further characterize and calibrate the MOI system. In a second step we investigate a heterostructure, where Co magnetic disks are positioned on top of a Nb superconducting film. Next, we consider the case of a micro-coil as a tunable inhomogeneous magnetic field source allowing us to precisely establish the field resolution of the MO indicator. Afterwards, we address a sample configuration where the whole superconducting Nb film is covered with a hard ferromagnetic NdFeB film with a predesigned chessboard pattern. Finally, we consider the case of a superconductor partially covered with a soft ferromagnetic permalloy (Py) layer, where the magnetic landscape can be customized by pre-imprinting a magnetic pattern into the Py layer.

\subsection{Arrays of magnetic bars}
\label{sec_bars}

\begin{figure*}[ht]
	\centering
	\includegraphics[width=\linewidth]{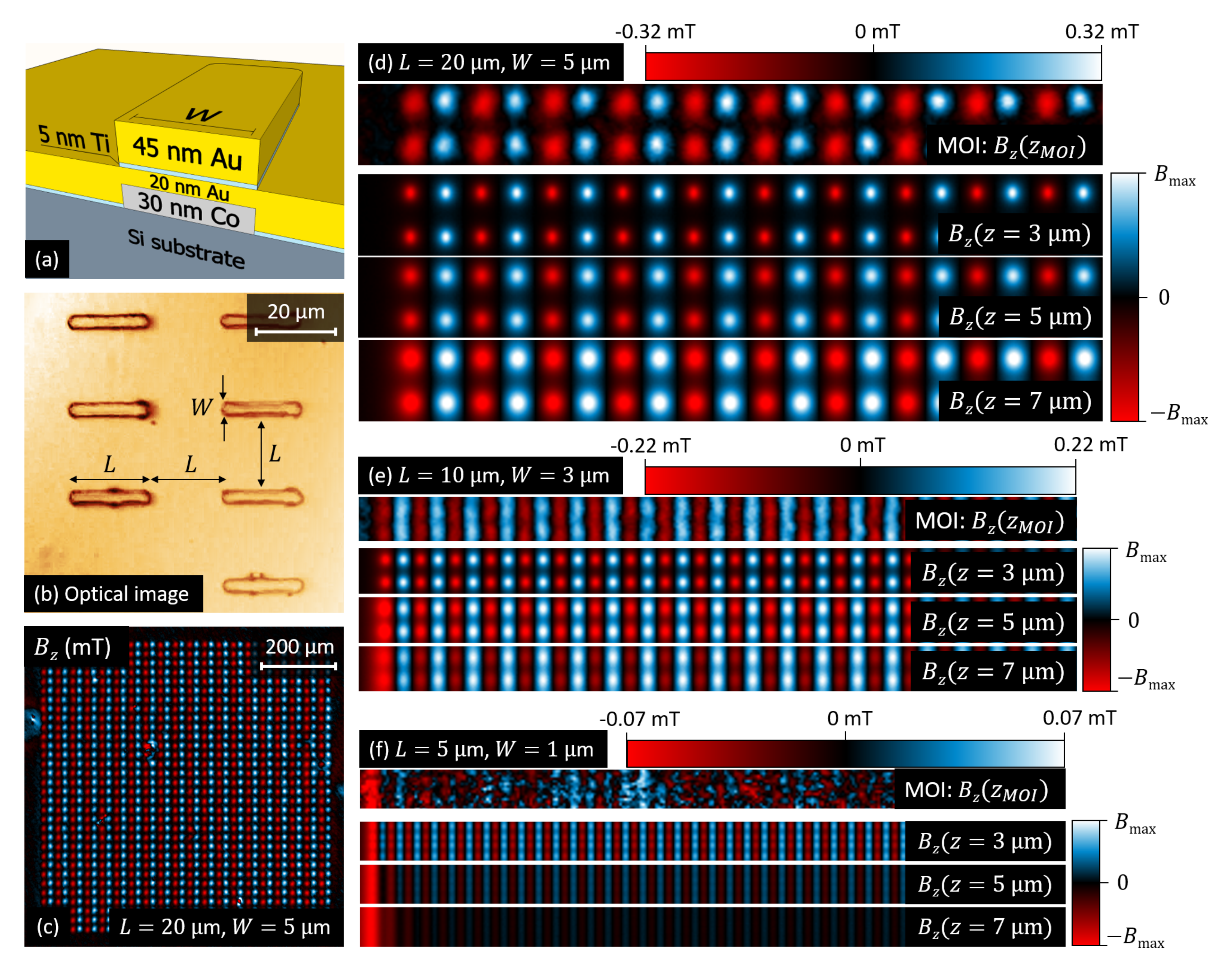}
	\caption{{\bf Quantitative magneto-optical imaging of magnetic bars.} (a) Schematic of a 30 nm thick Co magnetic bar of length $L$ and width $W$. (b) Optical image showing the periodicity of the Co bar arrays, characterized by a spacing $L$ between neighboring bars. (c) Magnetic field image, obtained via MOI, of a full $800 \times 800 \, \mu$m$^2$ array of magnetic bars with $L=20 \, \mu$m and $W=5 \, \mu$m. (d) Zoom on a half width of the array shown in panel (c). This field distribution $B_{z}(z_\mathrm{MOI})$ is compared with the exact analytic solution for $B_{z}(z)$, based on Eq. (\ref{eq:B_bar}), taking into account the MO active layer thickness $d\simeq 3 \, \mu$m. In panels (d-f), the leftmost red spots delineate the border of the array. (e) Same for a sample with $L=10 \, \mu$m and $W=3 \, \mu$m, at the limit where the stray field of a single bar can be resolved. (f) Same for a sample with $L=5 \, \mu$m and $W=1 \, \mu$m, where single bars cannot be seen any more and the response is dominated by the stray field at the border of the array. From the comparison of the theoretical and experimental magnetic field distributions, we estimate $z_\mathrm{MOI}$ to be between 5 $\mu$m and 7 $\mu$m.}
	\label{Fig-bars}
\end{figure*}

One of the crucial parameters determining the spatial resolution and the magnetic field reaching the magneto-optical indicator is the gap between the sample and the indicator placed on top of it. This distance is difficult to control, since it depends strongly on the roughness and the cleanliness of the sample and mirror surfaces, and it is therefore also challenging to reproduce, making it problematic to compare images from different experiments. Moreover, quantifying the spatial and field resolution of the technique is not straightforward, since it requires a precise knowledge of the field at the optically active layer position and the geometry of the field source. For these reasons, it is of interest to design localized magnetic field sources with a well-known field distribution. This can be achieved by using parallelepipedal magnetic bars, for which the magnetic field distribution follows an analytical expression\cite{Engel2005}.

Panels (a) and (b) of Fig. \ref{Fig-bars} show the structure of the sample, consisting of several $800 \times 800~\mu$m$^2$ arrays of nearly parallelepipedal 30 nm thick Co bars of length $L$ and width $W$. The samples are fabricated by a combination of e-beam lithography and MBE evaporation (cf. supplementary material for details). $L$ is the nominative length of the bars and does not take into account the round edges, effectively reducing the size of the domain with in-plane magnetization. All the arrays are fabricated on the same substrate, ensuring their observation under similar experimental conditions. We consider three different arrays: $L=20 \, \mu$m and $W=5 \, \mu$m, $L=10 \, \mu$m and $W=3 \, \mu$m, and $L=5 \, \mu$m and $W=1 \, \mu$m. 

We applied quantitative magneto-optical imaging to map the magnetic field $B_{z}(z_\mathrm{MOI})$ generated by the magnetic bars in the MO indicator, located at a distance $z_\mathrm{MOI}$ from the sample surface. As explained in the calibration procedure in section \ref{sec_calib}, we sweep the perpendicular applied magnetic field $\mu_0 H$ from +12.5 mT to -12.5 mT by steps of 0.1 mT and record the average of 3 images for every value of $H$. Before we start MOI, the Co bars are magnetized along their length with a permanent magnet. Note that the maximum applied field $\mu_0 H = 12.5$ mT does not change irreversibly the magnetization $M$ in the magnetic bars, a fact supported by the comparison of images before and after the calibration procedure. Any reversible changes in $M$ are accounted for by the calibration. Fig. \ref{Fig-bars}(c) shows the local magnetic field $B_{z}$ map of the magnetic bar array with $L=20 \, \mu$m and $W=5 \, \mu$m. In this image, each individual bar is identified by a pair of red and blue-white dots, representing the stray field of opposite polarity associated with the bar extremities. Moreover, the ratio $L/W=4$ is of the order of magnitude of the expected value for the appearance of single domains in Co bars\cite{Seynaeve2001}, a fact that is evidenced by the pairs of dots in the MO image. Our observation of well-defined poles at the ends of the bars agrees with either a mono-domain structure or at least with highly aligned domains. Since in our case the domain structure is driven by shape anisotropy, similar structures are also expected for the other two arrays.

Knowing the dimensions of the array of bars and approximating their geometry by a parallelepiped, we can calculate the local magnetic field map at a distance $z$ from the sample surface. Indeed, as was shown in Ref. \onlinecite{Engel2005}, the magnetic field $B(x,y,z)$ generated at the coordinates $(x,y,z)$ by a parallelepipedal single domain with magnetization $M$ and dimensions $L \times W \times t$ can be calculated analytically. The out-of-plane component of the magnetic field, $B_z(x,y,z)$, that MOI is sensitive to, can be expressed as follows:
\begin{widetext}
\begin{equation}\label{eq:B_bar}
\begin{split}
B_z (x,y,z) =& \frac{\mu_0 M}{4\pi} \sum_{k,l,m=1}^{2} (-1)^{k+l+m} \ln \biggl[ \left( x-x_0 \right) +(-1)^k \frac{W}{2} \\
&+ \sqrt{\left( \left( x-x_0 \right) +(-1)^k \frac{W}{2} \right)^2 
+ \left( \left( y-y_0 \right) +(-1)^l \frac{L}{2} \right)^2 
+ \left( z +(-1)^m \frac{t}{2} \right)^2 } \biggr].
\end{split}
\end{equation}
\end{widetext}
The bar has its center at the point of coordinates $(x_0,y_0,0)$ and has its main axis oriented along the $y$-axis. We add up the contributions of all the magnetic bars, obtained by changing $x_0$ and $y_0$ in the previous equation. Since the MO active layer of the indicator has a finite thickness $d=3 \, \mu$m, it is imperative to account for the fact that the magnetic field is not constant throughout the indicator. The MO images will therefore represent the average of the magnetic field over the distance $d$. In the calculations, the magnetic field distribution $B_{z}(x,y,z_0)$ obtained when the gap between the indicator and the sample surface is $z_0$ is obtained by averaging 31 magnetic field distributions $B_{z}(x,y,z)$, calculated by sweeping $z$ from $z_0$ to $z_0+d$ by steps of $0.1 \, \mu$m.

Figure \ref{Fig-bars}(d) shows a comparison of the experimental out-of-plane magnetic field $B_{z}(z_\mathrm{MOI})$ with the calculations based on Eq. (\ref{eq:B_bar}), for three distances $z_0$ between sample and indicator: 3 $\mu$m, 5 $\mu$m and 7 $\mu$m. The left side of the images corresponds to the left edge of the Co bar array shown in Fig. \ref{Fig-bars}(c). We chose linear color scales to represent the magnetic field $B_{z}$, meaning that we can directly compare visually the theoretical magnetic field distributions with the experimental one. This allows us to estimate the distance $z_\mathrm{MOI}$ between 5 $\mu$m and 7 $\mu$m. Moreover, the experimental $B_{z}$ image gives a maximum field $0.32\,\pm\,0.03$ mT. This can be compared with the maximum field $B_\mathrm{max}$ obtained with the analytical expression, where $M=1.4 \times 10^6$ A/m is taken as the saturation magnetization of Co\cite{Cullity2011}. This gives $B_\mathrm{max} = 0.99$ mT, $0.49$ mT and $0.29$ mT for distances $z_0$ of $3 \, \mu$m, $5 \, \mu$m and $7 \, \mu$m, respectively. These values allow for an estimation of $z_\mathrm{MOI} \sim 6.3 \, \mu$m falling in the range of the distances anticipated by visual inspection.

From the MO data, we observe that the distance between the two poles of a magnetic bar, i.e. the distance between the center of the red and blue-white dots, is not 20 $\mu$m as in the theoretical field maps, but is actually $\sim 18 \, \mu$m. This is due to the fact that the magnetic bars are not perfectly parallelepipedal but have round corners, as shown in the optical image in Fig. \ref{Fig-bars}(b). Therefore, the influence of the bars' borders, where the magnetization tends to align with the edges to minimize the energy of the system, reduces the effective size of the domain of magnetization $M$.

Panels (e) and (f) of Fig. \ref{Fig-bars} follow the same principle for arrays of magnetic bars with $L= 10 \, \mu$m and $W=3 \, \mu$m, and $L= 5 \, \mu$m and $W=1 \, \mu$m, respectively. In the MO-based field map of Fig. \ref{Fig-bars}(e), the single poles of the magnetic bars can barely be resolved, meaning the spatial resolution of our system is of the order of $5~\mu$m for this particular $z_\mathrm{MOI}$. The neighboring bars look nearly connected and form long stripes extending in the direction perpendicular to the bars main axis. Comparison with the theoretical field maps gives $z_\mathrm{MOI} \sim 6 \, \mu$m and confirms the value we found based on Fig. \ref{Fig-bars}(d). Note that here, in comparison to the bigger bars, the magnitude of the maximum magnetic field decayed to $0.22\,\pm\,0.03$ mT in the MO image, while it is $0.65$ mT, $0.30$ mT and $0.19$ mT in the calculations, for $z = 3 \, \mu$m, $5 \, \mu$m and $7 \, \mu$m, respectively.

In figure \ref{Fig-bars}(f), the perpendicular stripes observed in panel (e) nearly disappeared from the MO-based field map, leaving as the main feature in the image a bright line at the border of the bar array, where the field has a maximum magnitude of around $0.07\,\pm\,0.03$ mT. This means that the magnetic resolution of our system is better than 0.1 mT. The theoretical field maps give $B_\mathrm{max} = 0.26$ mT, $0.14$ mT and $0.09$ mT for $z = 3 \, \mu$m, $5 \, \mu$m and $7 \, \mu$m, respectively, which is in fair agreement with the $z_\mathrm{MOI}$ value estimated previously.

\subsection{Magnetic Co disks on top a superconducting Nb film}
\label{sec_dots}

\begin{figure*}[ht]
	\includegraphics[width=12.0cm]{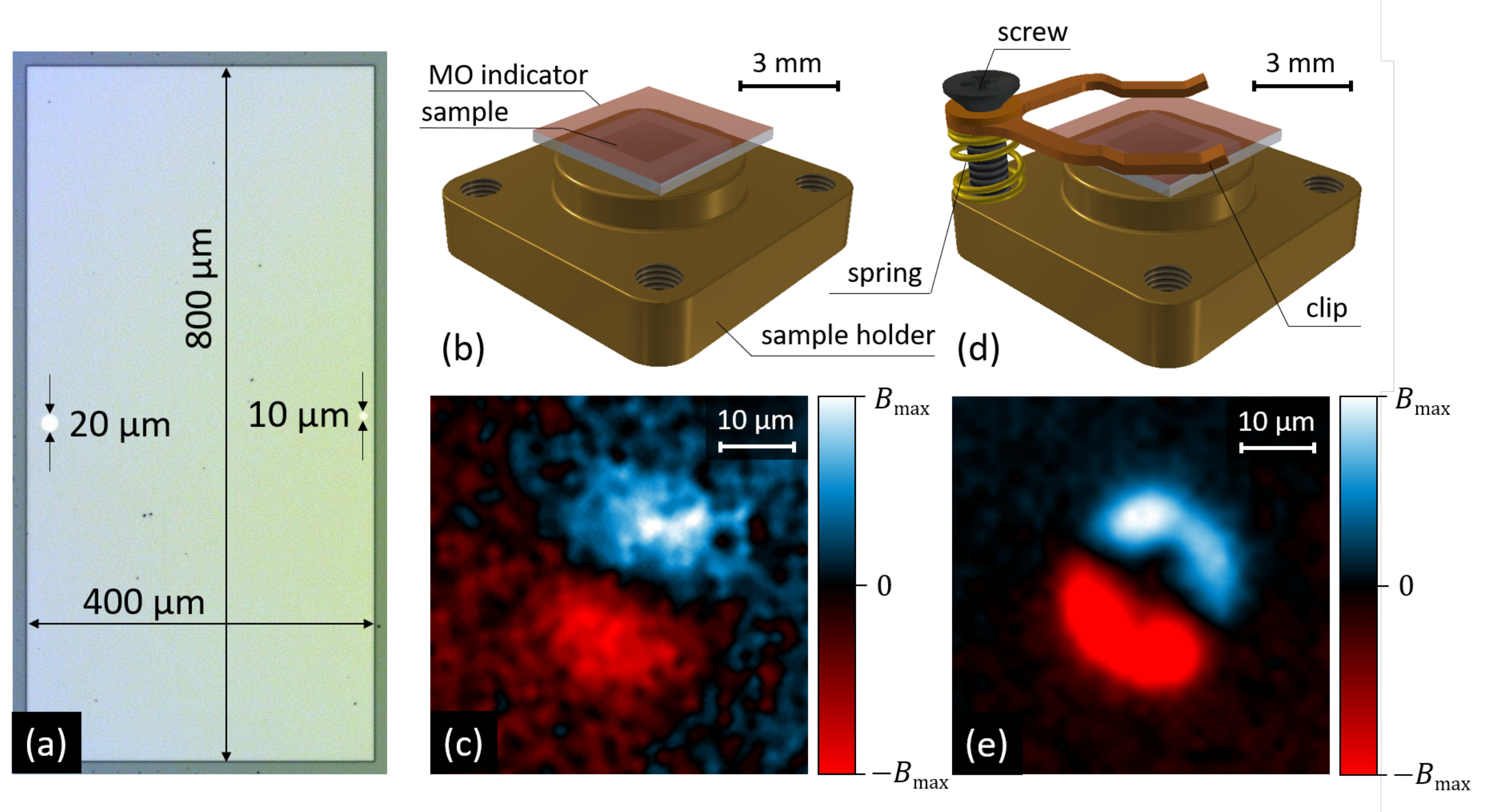}
	\caption{{\bf Sample layout and mounting for the Nb film with Co disks.} (a) Optical image of the $800 \times 400 \, \mu$m$^2$ Nb film (100 nm thick), where the two 30 nm thick Co disks with a diameter of $20 \, \mu$m and $10 \, \mu$m are located at $20 \, \mu$m and $10 \, \mu$m, respectively, from the sample edge. (b) Classical configuration for the MOI measurements, with the sample mounted on top of the sample holder, and the indicator placed on top of it. (c) MO image of the $20 \, \mu$m diameter Co disk after polarization with an in-plane field of 3 mT, in the configuration of panel (b). (d) Alternative configuration where the indicator is pressed closer to the sample surface with a purposely designed metallic clip. (e) MO image of the same disk as in panel (c) showing the enhanced MO signal using the alternative configuration presented in (d).}
	\label{Fig-dots-1}
\end{figure*}

\begin{figure*}[ht]
	\includegraphics[width=12.0cm]{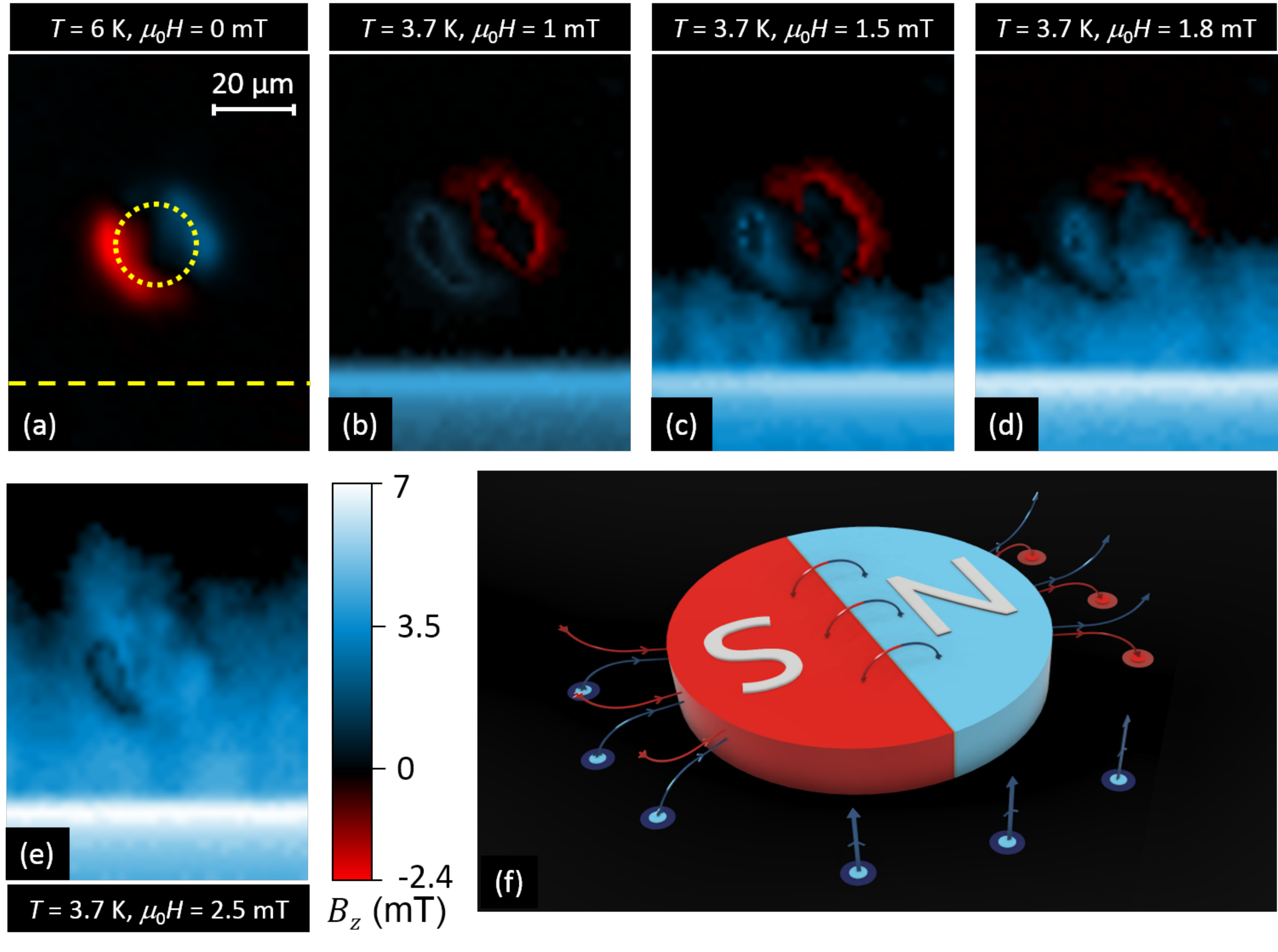}
	\caption{{\bf Influence of a Co disk on flux penetration in the superconducting Nb film.} (a) Magnetic field distribution of a $20 \, \mu$m diameter Co magnetic disk at $T=10$ K. The disk is outlined by the yellow dotted circle, while the border of the superconductor is marked by the dashed straight line. The sample is subsequently cooled down to $T=3.7$ K and a sequence of MO images is recorded for (b) $\mu_0 H=1$ mT, (c) $\mu_0 H=1.5$ mT, (d) $\mu_0 H=1.8$ mT and (e) $\mu_0 H=2.5$ mT. These images show the magnetic field distribution after the field of the disk, represented in (a), has been removed following the procedure described in section \ref{sec_calib}. Panel (f) shows a sketch of the interaction between a magnetized Co disk, assuming a perfect dipolar configuration, and the superconductor film below. The indicator positioned on top of it is not shown for the sake of clarity. Field lines representing negative (positive) $B_z$ components are in red (blue). Near the magnetized disk, the positive field as seen by the indicator is opposed to the field in the superconductor below. The vertical blue lines at the right side of (f) represent the field lines from the applied field $H$.}
	\label{Fig-dots-2}
\end{figure*}

In this section, we will increase the complexity of the MOI detected signal by adding a superconducting film to the localized source of magnetic field produced by magnetic disks. To that end, the very same system that has been used to illustrate the protocol for conversion from $I$ to $B_{z}$ in Fig. \ref{Fig-calib}, is now analyzed in more detail. The sample under study consists of a rectangular $800 \times 400 \, \mu$m$^2$ Nb film (100 nm thick) with two Co disks (30 nm thick) on top of its surface, as shown in Fig. \ref{Fig-dots-1}(a). The critical temperature of the Nb film is $T_\mathrm{c} = 9.0$ K, as determined by AC magnetic susceptibility measurements. DC measurements in a SQUID magnetometer confirmed a strong in-plane magnetic anisotropy of the Co disks.

Figure \ref{Fig-dots-1}(b) shows the classical MOI configuration used till this point in the paper. The MO indicator is placed on top of the sample, with a gap $z_\mathrm{MOI}$ between 5 and 7 $\mu$m between them, as determined in Section \ref{sec_bars}. We magnetized the disks with an in-plane field of the order of 3 mT using a commercial neodymium magnet. A MO image of the $20 \, \mu$m diameter Co disk at room temperature is shown in Fig. \ref{Fig-dots-1}(c). The stray field of the Co disk, indicating the direction of the in-plane magnetization, is visible in the image as a red (negative $B_z$) and blue-white (positive $B_z$) spot. Since the signal is quite weak, leading to poor contrast, it is important to minimize the gap between indicator and sample. The distance $z_\mathrm{MOI}$ can be considerably reduced if instead of just placing the MO indicator on the sample, a clip is used to press the indicator firmly onto the sample surface, as illustrated in the sketch of Fig. \ref{Fig-dots-1}(d). This leads to significant improvement in the MO contrast, as shown in the MO image in Fig. \ref{Fig-dots-1}(e), indicating a reduced effective distance $z_\mathrm{MOI}$ between indicator and sample surface. 

A negative side-effect of using the clip is that the mechanical stress induced by the clip favors the proliferation of in-plane magnetic domains in the indicator film. Since this effect severely affects the image quality, the use of a pressing clip is not systematic but rather limited to particular cases. In addition, sample mounting has been performed in an environment with a large amount of particles of at least 2.5 $\mu$m diameter in the atmosphere\cite{Air}. This restricts the reduction of the gap that can be achieved by pressing with a clip due to the unavoidable presence of such particles between the indicator and the sample.

The conversion procedure allowed us to determine the values of the magnetic field on top of and around the disks. At zero applied field, the extreme values around the poles are, $B_z = 2.4 \pm 0.3$ mT for the big disk (20 $\mu$m), and $B_z = 1.5 \pm 0.3$ mT for the $10 \, \mu$m disk. For the range of values of the applied field $H$, the field produced by the disk typically exceeds the field of the flux penetrating the superconducting sample, and therefore masks the contribution of the latter to the total field. For this reason, the image conversion method presented in section \ref{sec_calib} proves very useful, since it allows to isolate the contribution of the superconductor from the stronger signal of the Co disks.

In Figure \ref{Fig-dots-2}, we show the influence of the magnetic disks on flux penetration in the superconducting Nb film. Beforehand, the disks were magnetized at an angle with respect to the closest sample border, represented by the yellow dashed line in Fig. \ref{Fig-dots-2}(a). Panel (a) represents the magnetic field at $T=10$ K for the $20 \, \mu$m diameter disk, outlined with the dotted yellow circle, clearly visible as red and blue spots corresponding respectively to negative and positive $B_z$. The sample is subsequently cooled down to $T=3.7$ K and the applied field $H$ is increased. Images obtained following the procedure described in section \ref{sec_calib}, and where the magnetic field landscape in panel (a) is thus removed, show the magnetic field distribution for (b) $\mu_0 H=1$ mT, (c) $\mu_0 H =1.5$ mT, (d) $\mu_0 H=1.8$ mT, and (e) $\mu_0 H=2.5$ mT. In Fig. \ref{Fig-dots-2}(b), no magnetic field has penetrated into the superconducting layer yet and the magnetic disk signal has been reduced down to $0.1$ mT by carrying out the intensity-to-magnetic field conversion procedure. However, a weak magnetic signal is visible around the magnetic disk which might be attributed to the influence of the superconducting screening currents, repelling the magnetic field distribution of the disk and thus modifying slightly the field distribution shown in panel (a)\cite{Pokrovsky2006,Fraerman2005}.

When $H$ is increased, magnetic flux starts to penetrate into the sample in small flux jumps (Fig. \ref{Fig-dots-2}(c)) and interestingly, it can be clearly seen that it propagates preferentially through the side of the disk having the same polarity (positive $B_z$) as the applied field (Fig. \ref{Fig-dots-2}(d-e)). In order to understand this behavior, one must keep in mind that the polarity of the field generated by the disks in the vicinity of the poles is reversed in the indicator with respect to the superconductor\cite{Brisbois2016}, as shown in the sketch of \ref{Fig-dots-2}(f). The natural attraction (repulsion) between the vortices from the border and the antivortices (vortices) created by one of the poles is responsible for the asymmetry in flux penetration. This is visible in Fig. \ref{Fig-dots-2}(d-e) by the enhanced magnitude of the field at the preferred side of the disk, while the other side shows a perceptible shielding of flux. Similar results are obtained for the $10 \, \mu$m diameter magnetic disk, except that the influence on the flux penetration is weaker. Changing the orientation of the disk magnetization gives essentially the same results, the entering vortices being attracted (repelled) by the side of the disk with the same (opposed) polarity.


\subsection{Micro-scale electromagnet}
\label{sec_coil}

In the system described in Section \ref{sec_dots}, we have little control other than the direction of the magnetic moment of disks. An attempt to overcome this limitation is explored in this section where the magnetic disk has been replaced by a single turn micro-coil excited externally with a continuous current. This approach allows us to have a tunable yet localized source of inhomogeneous magnetic field essential for determining the magnetic field resolution of the technique. 

\begin{figure*}[ht]
	\includegraphics[width=14.0cm]{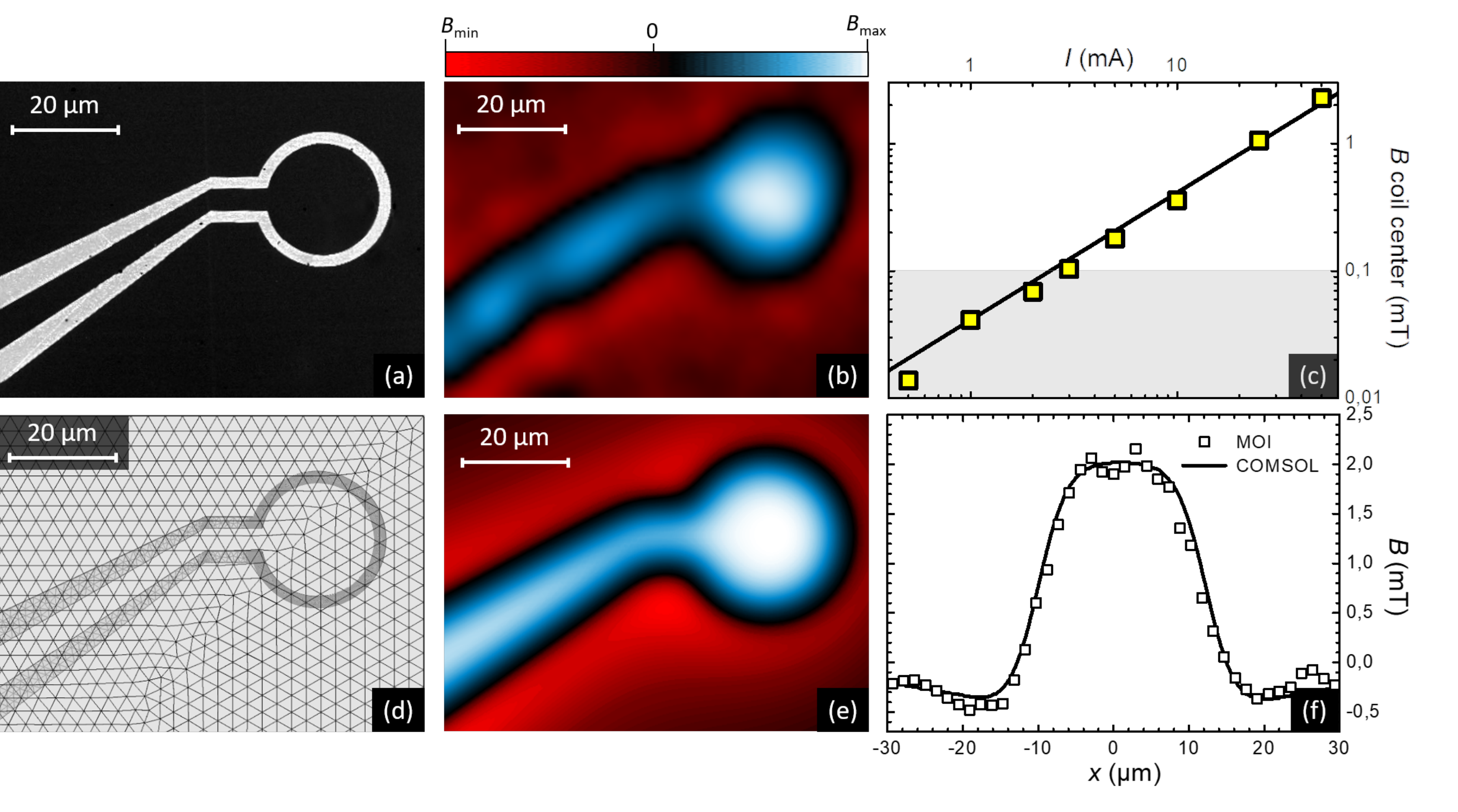}
	\caption{{\bf Quantitative magneto-optical imaging on a planar coil.} (a) Scanning electron microscope image of the 50 nm-thick coil made of Al. The coil has inner and outer radius of $10 \, \mu$m and $12 \, \mu$m, respectively. (b) Magnetic field of the coil fed by a current $I=50$ mA, obtained by MOI with the MO indicator pressed on the sample surface. (c) Magnetic field $B_z$ at the center of the coil as a function of $I$. The error on the magnetic field value of a single pixel is around $0.1$ mT, but for extended sources such as the coil, the correlation between pixels gives a lower detection threshold of $0.01$ mT. (d) Mesh for the numerical simulations. (e) Magnetic field of the coil fed by a current $I=50$ mA, obtained by numerical simulations, at a distance $z=6 \, \mu$m of the sample surface. (f) Magnetic field profile through the center of the coil for $I=50$ mA taken in the MO image of panel (b) and in the numerical simulation for $z=6 \, \mu$m.}
	\label{Fig-coil}
\end{figure*}

Figure \ref{Fig-coil}(a) shows a scanning electron microscope image of the sample layout. The Al coil is 50 nm thick and has inner and outer diameters of $20\, \mu$m and $24 \, \mu$m, respectively. All the measurements have been performed at room temperature. Figure \ref{Fig-coil}(b) shows a MO image of the out-of-plane magnetic field distribution of the coil for a current $I=50$ mA, obtained after applying the conversion procedure. The indicator was pressed on the sample using the clip mentioned in Section \ref{sec_dots}. To enhance the signal to noise ratio, the image shows the difference between the magnetic field obtained for positive and negative currents. 

Figure \ref{Fig-coil}(c) shows the out-of-plane magnetic field $B_z$ at the center of the coil as a function of the applied current amplitude $I$. For a coil of radius $R$ made of a unidimensional wire, $B_z$ on the coil axis at a distance $z$ from the coil center is given by:
\begin{equation}\label{eq:coil}
B_z = \frac{\mu_0 I R^2}{2 \left( R^2 + z^2 \right)^{3/2}}.
\end{equation}
Although the experimental coil has a finite thickness and its wire has a $2 \, \mu$m width, Eq. (\ref{eq:coil}) describes within an accuracy of 1\% the magnetic field of the real single-turn coil and can be used to fit the curve of Fig. \ref{Fig-coil}(c). The slope gives a value of $z_\mathrm{MOI} = 5.7 \pm 1 \, \mu$m. The error on the magnetic field for a single pixel is of the order of 0.1 mT, but the correlation between all the pixels forming the coil gives a significantly lower detection threshold of around 0.01 mT, at which the coil cannot be seen anymore. This value is in agreement with the resolution limit reported previously with similar MO indicators\cite{Koblischka1995} and could possibly be pushed further down by averaging a large number of images.

A more realistic description of magnetic field texture generated by the micro-coil can be obtained by numerical simulations, with the same geometry as the real coil. The mesh of finite elements used in the simulations is shown in Fig. \ref{Fig-coil}(d) and the magnetic field distribution obtained with the simulations for a current $I=50$ mA at a distance $z=6 \, \mu$m from the coil surface is represented in Fig. \ref{Fig-coil}(e). The good agreement between the experimental and the numerical results is further confirmed by plotting the magnetic field profile across the coil center for both the simulated images and the MO images, as shown in Fig. \ref{Fig-coil}(f). Note that the vertical distance $z=6 \, \mu$m obtained here corresponds approximately to the distance between the middle of the MO active layer and the coil, which gives a gap of about 4.5 $\mu$m between the coil and the bottom of the indicator.

\subsection{Nb/NdFeB heterostructures}
\label{sec_TMP}

Thus far we have tested the proposed protocol with prototypical inhomogeneous magnetic field sources of relatively weak intensity permitting us to determine the ultimate magnetic field resolution limit of the Bi:YIG MO indicator as 10 $\mu$T and a spatial resolution similar to the size of the gap separating the MO indicator and the source of inhomogeneous field. In what follows, we will exploit the quantitative MOI procedure in order to investigate more complex and innovative systems, with substantially higher magnetic fields across the sample area, and thus allowing us to test the proposed method in the opposite extreme of high magnetic fields and field gradients. It is worth emphasizing that to the best of our knowledge, application of extremely hard permanent magnetic materials, such as the NdFeB described hereunder as possible sources of vortex pinning remains relatively unexplored. These permanent magnetic materials are characterized by much higher coercive field $(\mu_\mathrm{0}H_\mathrm{c})$ and remanent magnetization $(\mu_\mathrm{0}M_\mathrm{r})$ than magnetic elements or binary alloys previously used in S/F hybrids. Indeed, for NdFeB\cite{Dempsey2007,Dumas-Bouchiat2010,Kustov2010} the reported value is $\mu_\mathrm{0}H_\mathrm{c}$ $\geq$ 1.5 T, much larger than those for Co/Pd multilayers ($\leq$ 0.6 T)\cite{Gillijns2005}, Co/Pt multilayers ($\leq$ 0.4 T)\cite{Gillijns2007}, Co dots ($\leq$ 0.03 T)\cite{Bael1999}, Co/Pd dots ($\leq$ 0.15 T)\cite{Lange2003}, Co/Pt dots ($\leq$ 0.23 T)\cite{VanBael2003,Neal2007,Gheorghe2008}, and Fe dots ($\leq$ 0.15 T)\cite{Villegas2007}, to name a few. Corresponding remanent magnetization values are also significantly larger for NdFeB in comparison to the other materials mentioned above. Hence, the use of magnetically patterned NdFeB as templates of spatially modulated magnetic field to control the pinning landscape in a superconductor may offer a number of advantages over softer magnetic materials, such as greater stability and larger magnetic field amplitudes of potential interest in technological applications.

\subsubsection{Characteristics of TMP NdFeB}
\label{subsec_TMP1}

\begin{figure*}[ht]
	\centering
	\includegraphics[width=12.0cm]{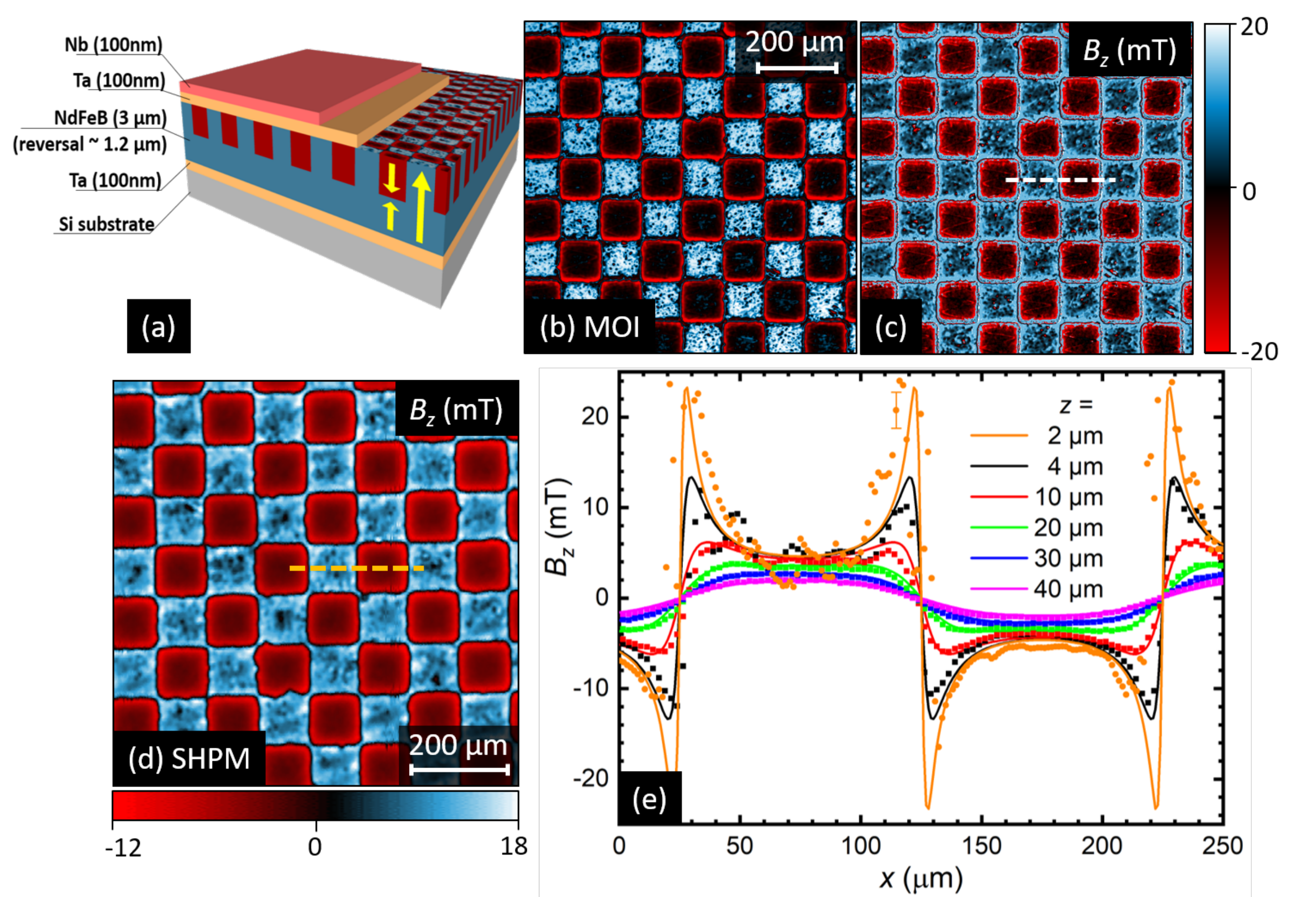}
	\caption{{\bf The Nb/NdFeB sample and characteristics of the TMP layer.} (a) Schematic of the Nb/NdFeB sample. (b) Part of a MO image of the Nb/NdFeB sample with chessboard pattern, obtained at $T$ = 10 K and $\mu_0 H$ = 1 mT. (c) $H_\mathrm{min}$ image showing the $B_{z}$ distribution in the chessboard pattern obtained by the calibration protocol (cf. text for details). (d) SHPM image of the Nb/NdFeB sample obtained at a scan height of $\sim$ 4 $\mu$m at room temperature showing a part of the chessboard pattern in NdFeB. (e) $B_{z}(x)$ profiles, from the calibration image in (c) (orange circles), and at different scan heights obtained from SHPM (squares), along the horizontal lines shown in (c) and (d). The profile from the calibration image is adjusted for the offset due to the non-zero analyzer angle. The error bar in one of the data points in this profile indicates the maximum extent of noise in the data (fluctuation in $B_{z}$ inside a square) $\sim \pm$ 2 mT. Also shown are $B_{z}(x)$ profiles at different heights from the sample surface obtained from calculations (cf. text for details).}
	\label{Fig-TMP-1}
\end{figure*}

Recently, a promising technique, thermomagnetic patterning (TMP), has been developed to produce magnetic fields spatially modulated in the range from tens to hundreds of microns\cite{Dumas-Bouchiat2010}. In this technique, the magnetization of a hard ferromagnetic film is initially saturated in one direction. The film is then irradiated by a pulsed laser through a mask, while an external field weaker than the film\textquoteright s room temperature value of coercive field is applied in the direction opposite to that of the magnetization of the film. Regions exposed to the irradiation are heated up, resulting in local reduction of coercivity and hence inducing reversal of magnetization in these regions. The final structure consists of an array of opposite magnetized micromagnets. Further details about the technique can be found in Ref. \onlinecite{Dumas-Bouchiat2010}. Patterning of a few microns thick high performance NdFeB hard magnetic films has been achieved using this technique. TMP can be used to produce a variety of configurations of micromagnets, e.g., chessboard, stripes, and periodic arrays of circular/square domains. We have prepared S/F hybrid structures by depositing a Nb film on top of such TMP NdFeB films. Figure \ref{Fig-TMP-1}(a) shows a schematic of a Nb/NdFeB sample showing the different layers. A 3 $\mu$m-thick NdFeB film is deposited on a Si wafer covered by a 100 nm-thick buffer layer of Ta. A capping layer of 100 nm-thick Ta is deposited on top of the NdFeB to avoid oxidation. A particular magnetic texture is imprinted in the NdFeB film by TMP and finally a 100 nm layer of Nb is deposited on top. The sample we primarily investigated is square-shaped (4 mm $\times$ 4 mm) and consists of a chessboard pattern of 100 $\times$ 100 $\mu$m$^2$ alternating squares with opposite magnetization in NdFeB (cf. arrows indicating the magnetization direction in adjoining squares in Fig. \ref{Fig-TMP-1}(a)). 

The relatively large size of the patterned magnetic domains in NdFeB can be revealed by MOI carried out above the $T_{c}$ of Nb. Figure \ref{Fig-TMP-1}(b) shows a MO image of the sample obtained at $T$ = 10 K and $\mu_0 H$ = 1 mT whereas Fig. \ref{Fig-TMP-1}(c) shows a $H_\mathrm{min}$ image generated using the calibration protocol described in Section \ref{sec_calib}. The $H_\mathrm{min}$ image essentially shows the $B_{z}$ distribution in the chessboard TMP pattern, with an offset due to the non-zero analyzer angle used in our measurements. The calibration works well to reveal the relatively large magnetic fields in the chessboard pattern, except at the borders of the oppositely magnetized squares, where many of the pixels are saturated due to large magnetic fields in these locations. The contrast of the image in Fig. \ref{Fig-TMP-1}(c) is adjusted to mask the saturating effect of such pixels.

Magnetic properties of the NdFeB layer have also been studied using SHPM performed at room temperature\cite{Shaw2016}. SHPM images were obtained at various scan heights from the sample surface. Figure \ref{Fig-TMP-1}(d) shows a SHPM image of the sample showing the chessboard pattern in the NdFeB obtained at the closest scan height $\sim$ 4 $\mu$m. From both the MO and SHPM images it is clear that the magnetic landscape inside the two sets of squares in the pattern are distinct from each other. While the \textquoteleft red\textquoteright\space squares in the SHPM image appear quite smooth, the \textquoteleft blue\textquoteright\space squares appear rather coarse in comparison. This is a result of the TMP process. Since magnetization reversal induced by the laser occurs at an elevated temperature and the external field applied during TMP is relatively weak ($\sim$ 0.5 T), the magnetic profile in the reversed (irradiated) regions is coarser in comparison to that in the non-reversed (non-irradiated) regions.

A set of $B_{z}(x)$ profiles, one from the MOI calibration image in Fig. \ref{Fig-TMP-1}(c) (orange circles) and the rest from SHPM images (squares) at different scan heights along the dashed lines shown in panels (c) and (d), are shown in panel (e). Note that a few data points, corresponding to saturated pixels at the square boundaries, have been removed from the MOI profile. We have also computed the $B_{z}$ distribution in the chessboard pattern (a 40 $\times$ 40 array of 100 $\times$ 100 $\mu$m$^2$ squares) using the analytical approach discussed for an array of magnetic bars in Section \ref{sec_bars}. The resulting $B_{z}(x)$ profiles (colored lines) at various heights from the sample surface are shown together with the experimental profiles in Fig. \ref{Fig-TMP-1}(e). For simplicity, a fixed value of remnant magnetization\cite{Kustov2010} $M$ = $2.8 \times 10^{5}$ A/m was assumed for the whole sample such that the calculated profiles at heights 4-40 $\mu$m agree well with the experimental profiles obtained using SHPM. The calculated profile thus obtained at 2 $\mu$m agrees reasonably well with the profile obtained from MOI. This indicates that the effective height of the MO imaging plane from the sample surface ($z_\mathrm{MOI}$) is in this case $\sim$ 2 $\mu$m, which is expected as the indicator was pressed with the clip for this set of measurements. The estimations above do not take into account the fact that Nd\textsubscript{2}Fe\textsubscript{14}B undergoes a spin reorientation transition from easy-axis to easy-cone configuration, with a cone angle of $30^{\circ}$, at 135 K. This would lead to a reduction of the magnetization by about 13\%, in addition to the effects produced by a non-perfectly $c$-axis oriented magnetic texturing\cite{Givord1985,Garcia2000}. It is worth noting that the $B_{z}$ profiles in the TMP pattern resemble that expected for a ferromagnetic film with domain widths much larger than the film thickness\cite{Aladyshkin2003,Aladyshkin2009}, with the magnetic field decaying from a domain wall towards its center.

\subsubsection{Isothermal magnetic flux penetration in Nb/NdFeB hybrid system}
\label{subsec_TMP2}

\begin{figure}[ht]
	\centering
	\includegraphics[width=8.5cm]{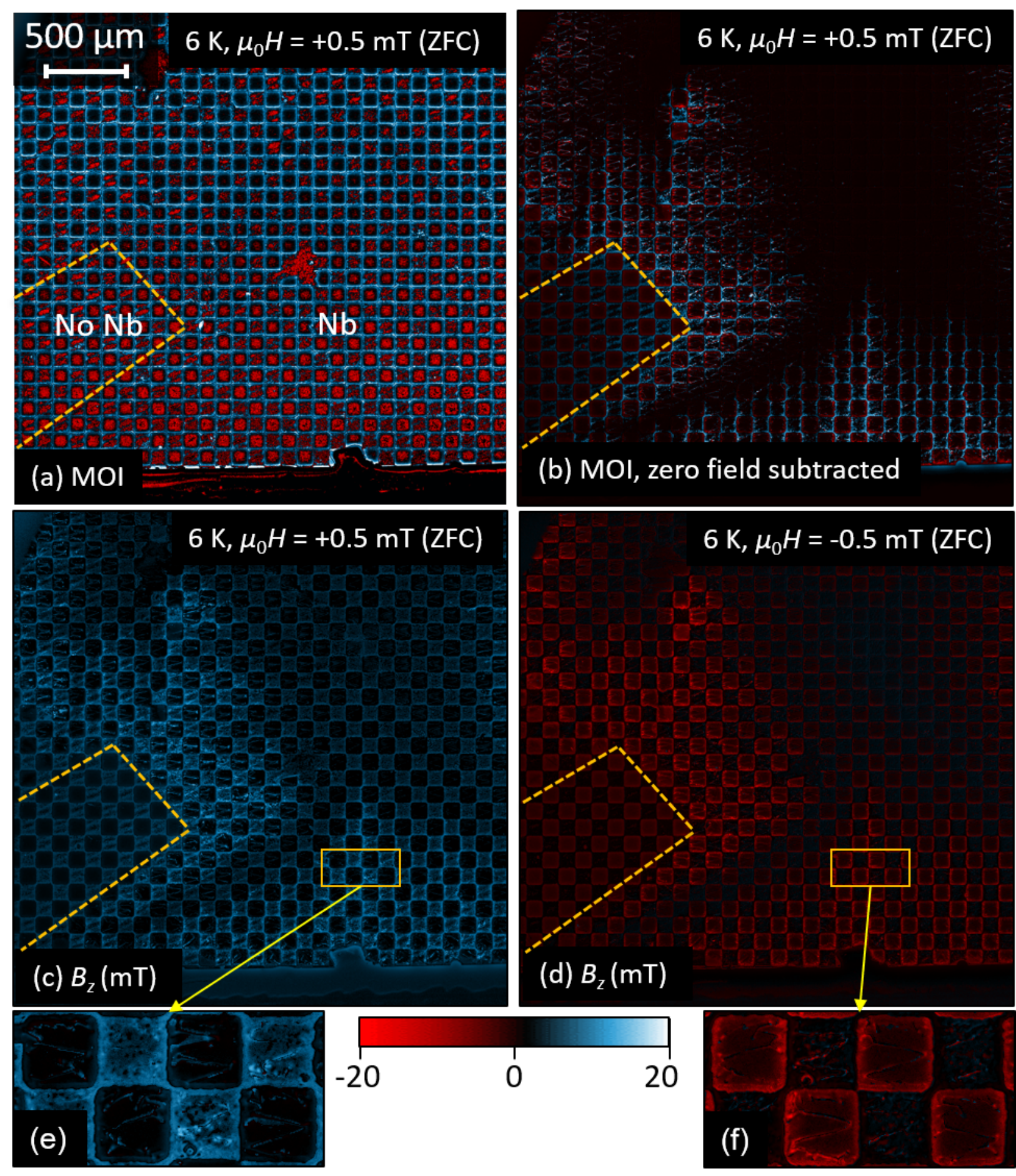}
	\caption{{\bf Smooth flux penetration in Nb/NdFeB at 6 K.} (a) MO Image obtained at +0.5 mT while increasing $\mu_0 H$ from 0 mT after zero-field cooling the sample to 6 K. A region without Nb at lower-left of the image is indicated by the dashed lines. The image in (b) is obtained by subtracting the MO image at 6 K and 0 mT (i.e., before applying any magnetic field after cooling down to 6 K) from the MO image shown in (a). (c) Image of $B_{z}$ distribution converted from the MO image shown in (a) using the calibration procedure. (d) Image of $B_{z}$ distribution converted from MO image obtained at -0.5 mT while decreasing $\mu_0 H$ from 0 mT after zero-field cooling. (e)-(f) Images cropped and zoomed from (c) and (d), respectively, showing the flux distribution in a few squares for the two scenarios. The $B_{z}$ distribution in (c)-(f) is indicated by the adjoining scale bar.}
	\label{Fig-TMP-2}
\end{figure}

In order to visualize magnetic flux penetration in Nb, the sample was cooled down to 6 K $< T_\mathrm{c}$ in zero field (ZFC). Then a series of MO images were obtained by cycling $\mu_0 H: 0 \rightarrow +12.5 \rightarrow -12.5 \rightarrow 0$ mT. Figure \ref{Fig-TMP-2}(a) shows a raw MO image obtained in the first run at $\mu_0 H$ = +0.5 mT (increasing from $\mu_0 H$ = 0 mT). A region without Nb, a consequence of the shadow of a holding clip during Nb deposition, is demarcated with dashed lines near the lower left of the image. In this figure, no hint of a superconducting response can be seen due to the overwhelming signal from the underlying NdFeB concealing the much weaker signal from Nb. A possible way to recover the superconducting signal from Nb is to take the difference of this image with the image obtained at 6 K and zero magnetic field, as shown in Figure \ref{Fig-TMP-2}(b). In this difference image, quite a few features are revealed. The region without Nb is readily identified. In the region with Nb, the smooth magnetic flux penetration into the Nb film progressively unveils the chessboard pattern underneath. The region of dark contrast towards the upper right corner of the image is where magnetic flux is absent. 


Let us now compare this approach with the result of applying the protocol for converting the intensity images into $B_{z}$ maps. Figure \ref{Fig-TMP-2}(c) shows the resulting $B_{z}$ map converted from the MO image in panel (a). From this figure it can be seen that our protocol is able to recover the smooth magnetic flux penetration into the Nb film. However, the fact that the chessboard pattern remains faintly visible in the Meissner region indicates that it fails to properly correct the magnetic field of the underlying magnetic pattern. This is understandable since the MO indicator saturates and therefore becomes insensitive to fields above 80 mT whereas the magnetic field close to the boundary between two squares can largely exceed that value. 

A closer inspection of the box in Fig. \ref{Fig-TMP-2}(c), and the zoomed-in view of this region shown in Fig. \ref{Fig-TMP-2}(e), reveals that the red squares are devoid of flux, indicated by the uniform dark contrast whereas a strong accumulation of flux is seen in the peripheral regions of the blue squares. When $(H)$ polarity is reversed, the opposite effect is observed as shown in Fig. \ref{Fig-TMP-2}(d). In this case, flux accumulation is observed inside the red squares (cf. strong red contrast in the periphery of the red squares in Figs. \ref{Fig-TMP-2}(d) and \ref{Fig-TMP-2}(f)). In other words, the flux propagates following staircase-like paths along one set of squares, while largely avoiding the other set. 

Vortices generated in a superconductor by an external magnetic field experience magnetic pinning by a domain structure due to interactions of the stray fields of the magnetic structure with screening currents in the superconductor\cite{Bulaevskii2000,Erdin2001,Bespyatykh2001,Milo2002,Laiho2003,Aladyshkin2009}. Moreover, the magnetic pattern is expected to spontaneously induce vortices in the superconductor even in zero applied field, with vortices of opposite polarity occupying the alternate sets of squares\cite{Erdin2001,Laiho2003,Aladyshkin2006}. In this scenario, a vortex generated in Nb by a positive applied $H$ approaching a red square will be attracted and annihilated by antivortices present at the edges of the square. While approaching the edge of a blue square, it will encounter a permeable wall of repulsive vortices, hence flux entry into the blue square is not impeded. Once inside a blue square, the vortex will be strongly pinned by virtue of magnetic pinning within the domain. This would lead to an accumulation of flux inside the blue squares. In agreement with the variation of $B_{z}$ within a square, more vortices would tend to accumulate near the edges of the square rather than near its center, due to the larger $B_{z}$ near the edge, as is observed in Figs. \ref{Fig-TMP-2}(c) and \ref{Fig-TMP-2}(e). For a negative applied $H$, the scenario will be reversed, with vortices accumulating in the red squares (cf. Figs. \ref{Fig-TMP-2}(d) and \ref{Fig-TMP-2}(f)). This is consistent with earlier reports demonstrating spatial variation of flux density in a superconductor guided by the magnetic landscape in its vicinity (e.g, \onlinecite{Laviano2005,Gillijns2007,Iavarone2014}), and will progressively result in a staircase-like path of flux propagation with increasing $H$.

\subsubsection{Magnetic flux avalanches in Nb/NdFeB hybrid system}
\label{subsec_TMP3}

\begin{figure}[ht]
	\centering
	\includegraphics[width=8.5cm]{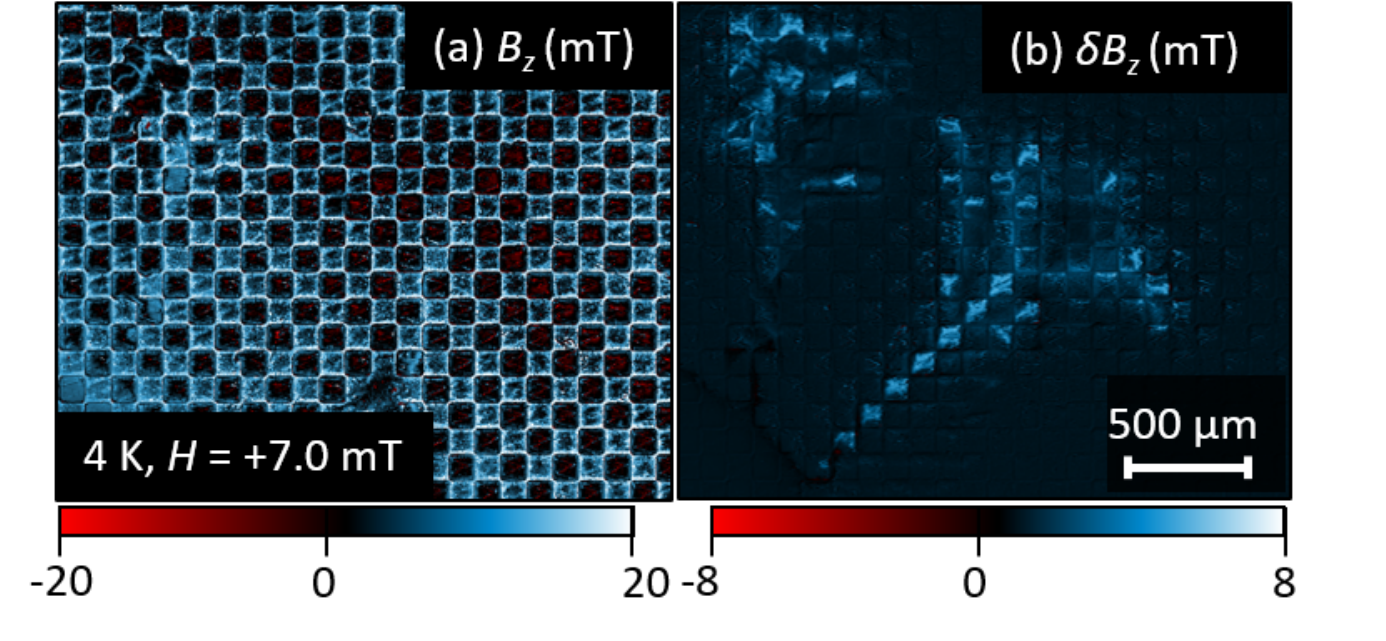}
	\caption{{\bf Flux avalanches in Nb/NdFeB at 4 K.} (a) Image of $B_{z}$ distribution converted from MO image obtained at 7 mT, while increasing $\mu_0 H$ from 0 mT after zero-field cooling the sample to 4 K. (b) Differential image showing $\delta B_{z}$ map in response to a 0.1 mT change in $\mu_0 H$ at 7 mT.}
	\label{Fig-TMP-3}
\end{figure}

It is a well documented fact that the increase of critical current density along with the decrease of thermal conductivity as temperature decreases, lead to the development of thermomagnetic instabilities\cite{Mints1981}. These instabilities manifest themselves as a burst of magnetic flux penetrating into the superconductor and acquire a particularly dramatic aspect in thin films with a branching structure similar to Lichtenberg figures\cite{Takahashi1979}. Literature on the influence of a magnetic template on the propagation of magnetic flux avalanches is rather scarce\cite{Gheorghe2008,Brisbois2016} and motivates us to explore this regime of flux penetration in the Nb/NdFeB hybrid system. To that end, the sample is cooled down to the base temperature of the cryostat ($\sim$ 4 K) in zero field (ZFC). Figure \ref{Fig-TMP-3} summarizes our observations of avalanche-like flux jumps observed in this regime. Fig. \ref{Fig-TMP-3}(a) shows the magnetic field map for $\mu_0 H$ = 7 mT where the Nb sample has been fully penetrated and no evidence of avalanche is seen. However, by taking the differential image ${\delta}B_{z} = B_{z}(\mu_0 H+0.1$~mT)$-B_{z}(\mu_0 H)$ as shown in Fig. \ref{Fig-TMP-3}(b), a sudden large jump is observed along a diagonal, in an area encompassing a set of squares of similar magnetization. This is somewhat analogous to the observation of secondary branches of flux avalanches in Nb and MoGe films with an antidot lattice\cite{Motta2014}. By further increasing $H$ a smooth flux penetration proceeds, which is interspersed by similar jumps at higher $H$. 

It is worth noting that the flux jumps observed at 4 K occur over and above smooth flux penetration. This observation leads us to suggest that the observed avalanches might not be the result of thermomagnetic instabilities but rather the effect of flux channeling. Further experimental investigations will be needed to clarify the origin of the flux jumps reported here.

\subsection{Py/Nb heterostructures}
\label{sec_NbPy}

\begin{figure*}[ht]
	\centering
	\includegraphics[width=15.0cm]{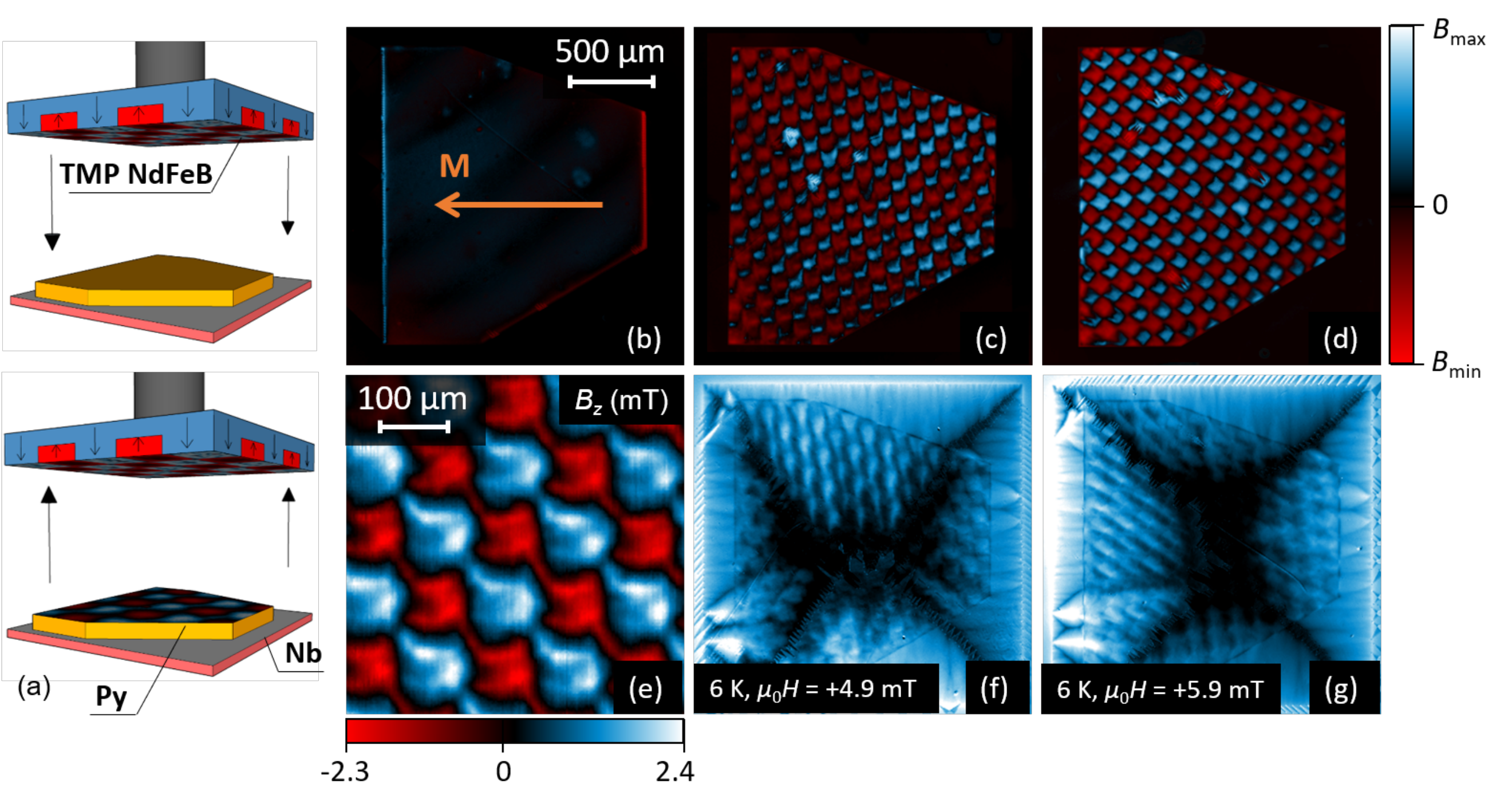}
	\caption{{\bf Imprinting the micromagnetic pattern in permalloy.} (a) Schematic showing the imprinting process. (b) Py/Nb sample magnetized in-plane with the magnetization direction indicated by the arrow. (c)-(d) Py/Nb sample with micromagnetic pattern imprinted in the Py layer. (c) Chessboard pattern aligned roughly along the long side of Py. (d) Chessboard pattern at $\sim 45^{\circ}$ with respect to the long side. (e) SHPM image at room temperature showing a part of the chessboard pattern in the Py layer of panel (c), obtained at a scan height $z \sim$ 4 $\mu$m. (f)-(g) MO Images showing smooth flux penetration at 6 K in the Py/Nb sample with the imprinted patterns shown in (c) and (d), respectively. These images are obtained by subtracting the image at $\mu_0 H$ = 0 mT in each set to remove the signal from the magnetic pattern. Images in (f) and (g) are at $\mu_0 H$ = 4.9 mT and 5.9 mT, respectively. Both images were obtained while increasing $\mu_0 H$ from 0 mT after zero-field cooling the sample to 6 K.}
	\label{Fig-NbPy-1}
\end{figure*}

In this section we will address the possibility to reproduce the magnetic landscape of TMP templates in a softer ferromagnet, so as to obtain a weaker, erasable, and tailor-made pinning potential.

Permalloy (Py) is an interesting material to use as a source of flexible magnetic landscape to induce vortex pinning in a superconductor\cite{Brisbois2016}. This can be achieved by transferring the TMP templates in NdFeB onto a 460 nm thick Py layer partially covering a Nb film. The imprinting process starts from a state where the Py layer is magnetized in-plane along the direction indicated in Fig. \ref{Fig-NbPy-1}(b). Then, the Py/Nb sample is fixed at the base of a micro-manipulator probe station with the Py layer on top and the base is clamped (cf. schematic in Fig. \ref{Fig-NbPy-1}(a)). The NdFeB sample is attached to a probe, with the patterned NdFeB surface facing down, towards the Py/Nb sample. Then the probe with the NdFeB sample is approached towards the Py/Nb sample till the two surfaces are in contact. Afterwards, the base is unclamped and pulled down ensuring minimal slipping between the two surfaces which results in a stable and clear imprinting of the TMP pattern in the Py. The TMP chessboard pattern was used to generate two different configurations in Py as shown in Figs. \ref{Fig-NbPy-1}(c) and \ref{Fig-NbPy-1}(d), with the chessboard pattern along and at $45^\circ$ with respect to the long edge of Py, respectively.

In order to expose the small scale details of the imprinted magnetic landscape, the Py layer of Fig.~\ref{Fig-NbPy-1}(d) has been characterized using SHPM performed at room temperature. The SHPM image in Fig. \ref{Fig-NbPy-1}(e) shows the chessboard pattern in the Py obtained at the closest scan height $z \sim$ 4 $\mu$m. $B_{z}$ in the pattern varies by 5 mT, which is $\sim$ 6 times less than that observed in NdFeB (cf. Fig. \ref{Fig-TMP-1}(d)), consistent with the fact that Py has a much lower remanent field than NdFeB. 

As with the Nb/NdFeB sample, to visualize flux profiles in Nb, the sample was cooled down to 6 K in zero field (ZFC). Then a series of MO images were obtained by cycling $\mu_0 H: 0 \rightarrow +12.5 \rightarrow -12.5 \rightarrow 0$ mT. These measurements were performed for both of the imprinted configurations shown in Fig. \ref{Fig-NbPy-1}. Panels (f) and (g) show one MO image for each configuration obtained in these sets. These images are obtained by subtracting the image at $\mu_0 H$ = 0 mT in each set to remove the signal from the magnetic pattern. The images in Figs. \ref{Fig-NbPy-1}(f) and \ref{Fig-NbPy-1}(g) are at $\mu_0 H$ = 4.9 mT and 5.9 mT, respectively. From these images, the modulation of flux path in Nb by the magnetic pattern is clearly established. The staircase-like paths of flux flow, which was also observed in the Nb/NdFeB sample, is much clearer in this sample. The principal reason behind this is that the magnetic signal from Py is much weaker than that from NdFeB, which allows revealing the response of Nb more easily. Furthermore, similar to our observations in Ref. \onlinecite{Brisbois2016}, flux motion is observed to be asymmetric with respect to the different edges of Py. Flux in Nb guided by the chessboard pattern penetrates much more quickly along one side of sample with respect to the others (cf. top of the sample in Fig. \ref{Fig-NbPy-1}(f) and left side of the sample in Fig. \ref{Fig-NbPy-1}(g)). This indicates that the underlying in-plane magnetization might still play a significant role even after imprinting the out-of-plane magnetic pattern in Py. A preliminary investigation of the flux jumps in this system shows that large dendritic avalanches develop at low temperatures, most probably associated with thermomagnetic instabilities.

\section{Conclusion}
\label{sec_conclusion}

In summary, we have developed a comprehensive protocol for calibration and conversion of MOI data into magnetic field distribution. A side benefit of this method is the removal of unwanted experimental artifacts from the raw data. The protocols have been applied on systems with increasing complexity. Several low magnetic field sources have been used to determine the magnetic field resolution of the technique as 10 $\mu$T whereas the spatial resolution has been shown to be similar to the gap separating the magnetic source and the MO indicator. For typical mounting conditions this spatial resolution lies between 2 and 10 $\mu$m. The introduced protocol helps to extract the comparatively weaker magnetic response of the superconductor from the background of larger fields associated with the magnetic layer in its vicinity. This has been notably useful to reveal magnetic flux penetration in Nb/NdFeB hybrids with a chessboard magnetic pattern. For this system, smooth flux penetration in Nb is observed to be strongly influenced by the underlying micromagnetic pattern, with incoming vortices preferentially occupying one set of squares of the pattern. Smooth flux penetration at lower $T$ is interspersed with unconventional avalanche-like flux jumps. In addition, thermomagnetically patterned micromagnet structures have been imprinted in permalloy (Py) to obtain flexible magnetic landscapes for flux guidance in a Nb layer underneath. Further refinements of the technique could be envisaged by incorporating deconvolution tools and taking into account the in-plane field component in the MO indicator.

\section{Supplementary Material}
\label{sec_supplementary}
See supplementary material below for more details on our magneto-optical imaging setup and on the preparation of the Co bars discussed in Section \ref{sec_bars}.

\begin{acknowledgments} This work was partially supported by the Fonds de la Recherche Scientifique - FNRS, the ARC grant 13/18-08 for Concerted Research Actions, financed by the French Community of Belgium (Wallonia-Brussels Federation), the COST action NanoCoHybri (CA16218), the Brazilian National Council for Scientific and Technological Development (CNPq) and the Sao Paulo Research Foundation (FAPESP). J.B. acknowledges support from F.R.S.-FNRS (Research Fellowship). The work of G.S. is supported by the University of Li\`{e}ge and the EU in the context of the FP7-PEOPLE-COFUND-BeIPD project. The work of S.B.A., A.V.S., and S.M. is partially supported by PDR T.0106.16 of the F.R.S.-FNRS. The authors thank the ULg Microscopy facility CAREM for part of the SEM investigations. L.B.L.G.P. was supported by a fellowship from CNPq-CsF Program. The LANEF framework (ANR-10-LABX-51-01) and the Nanoscience Foundation are acknowledged for their support with mutualized infrastructure.
\end{acknowledgments}

%


\begin{thebibliography}{71}
\bibitem [{shi()}]{shift}%
	\BibitemOpen
	\href@noop {} {}\bibinfo {howpublished} {The constant shift in intensity is
		mainly caused by the Voigt effect and the longitudinal Faraday effect. While
		the first is caused by linear magnetic birefrigence and takes place even for
		a light beam at normal incidence, the latter occurs when the incident light
		is not perfectly perpendicular to the indicator surface. Since the light has
		a small in-plane component, some unwanted rotation of polarization occurs
		depending on the direction of M in the indicator. This comes mainly from the
		illumination path of the microscope, but can also arise from the difficulty
		to mount the indicator perfectly horizontally.}\BibitemShut {Stop}%
\bibitem [{Air()}]{Air}%
	\BibitemOpen
	\href@noop {} {}\bibinfo {howpublished} {We estimated the air quality in our
		lab using a commercial particle monitor (Dylos DC1100-PRO PC) which revealed
		the presence of particles of at least 2.5 $\mu$m diameter in the
		air.}\BibitemShut {Stop}%
\end{thebibliography}

%

\pagebreak
\makeatletter
\onecolumngrid

\begin{center}
\textbf{\Large Supplementary material: Quantitative magneto-optical investigation of superconductor/ferromagnet hybrid structures}
\end{center}


\setcounter{equation}{0}
\setcounter{figure}{0}
\setcounter{table}{0}
\setcounter{section}{0}

\renewcommand{\theequation}{S\arabic{equation}}
\renewcommand{\thefigure}{S\arabic{figure}}
\renewcommand{\thesection}{S\arabic{section}}
\renewcommand{\bibnumfmt}[1]{[S#1]}
\renewcommand{\citenumfont}[1]{S#1}

\begin{figure*}[ht]
	\centering
	\includegraphics[width=14.0cm]{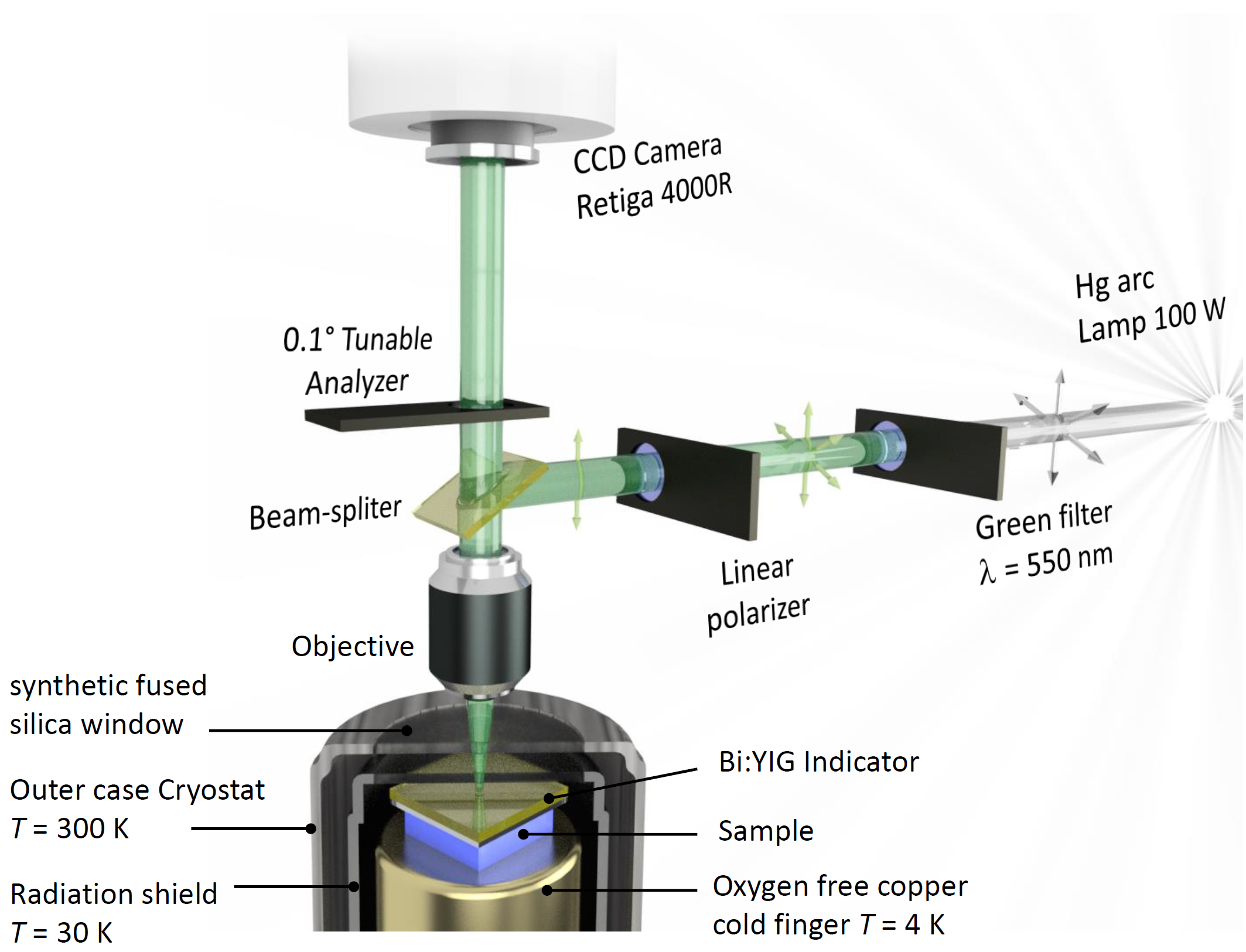}
	\caption{Schematic of the MOI setup (cf. text for details).}
	\label{Fig-MOI}
\end{figure*}

\twocolumngrid

\section{The Magneto-optical imaging setup}
\label{sec_MOI}

Magneto-optical imaging (MOI) is based on the Faraday effect, the rotation of the direction of polarization of a light beam proportional to the local magnetic field. Figure \ref{Fig-MOI} shows a schematic representation of the MOI setup at the University of Li\`{e}ge. The polarization microscope is a commercial Olympus modular system. The core of the microscope is the modular Olympus BX-RLA2 illuminator. Light beam produced by a 100 W Hg arc burner lamp (USH 103 D) is passed through a green filter (U-25IF550 at 550 nm), then crosses a linear polarizer (U-PO3), and is then directed to the Faraday active indicator through the objective by a beam splitter. As discussed in the manuscript, the indicator we use throughout this work is a 3~$\mu$m thick Bi-doped yttrium iron garnet (Bi:YIG) epitaxially grown on a 450~$\mu$m thick Gd$_3$Ga$_5$O$_{12}$ (GGG) transparent substrate. The Bi:YIG Faraday-active layer of the indicator has a Verdet constant $V = 0.018 \pm 0.005°$~$\mu$m$^{-1}$mT$^{-1}$ at 10~K. Typical values of the out-of-plane saturation field for indicators similar to ours are $\mu_0 H \sim 100$~mT. A 100~nm thick Al mirror was deposited on the optically active layer side in order to ensure sufficient reflection of the incident light beam. The linearly polarized light crosses the GGG substrate of the indicator and the Bi:YIG layer, where its polarization direction is rotated proportionally to the local magnetic field. It is then reflected by the mirror and crosses the indicator and the objective once again. It then passes through an analyzer (AN-360), whose polarization direction is oriented close to $90^{\circ}$ with respect to the polarizer. The orientation of the analyzer can be adjusted with a precision of $0.1^{\circ}$. The analyzer absorbs the component of light polarized in the original direction, so only the light whose polarization has been rotated in the indicator passes through. The rotation of the polarization is proportional to the component of the magnetic moment along the direction of light propagation. Light finally enters a high resolution RETIGA-4000R CCD-camera mounted on top of the microscope unit. The captor consists of 4.2 mega pixels and each pixel is 7.4 $\mu$m $\times$ 7.4 $\mu$m large. The camera captures light intensity on a 12-bit gray-scale and records $2048 \times 2048$ px$^{2}$ images where intensity values range from 0 to 4095. With a $5\times$ objective (LMPLFLN 5BD) this represents a field of view of approximately $3 \times 3$ mm$^{2}$, i.e., each pixel in the images corresponds to an area of 1.468 $\times$ 1.468 $\mu$m$^{2}$. We thus obtain a light intensity map representative of the magnetic field texture at the indicator's plane, where dark areas correspond to low magnetic fields and bright regions represent high fields. Magnetic fields in the range $\pm 12.5$ mT are applied by feeding a cylindrical copper coil with a dc current. The sample is installed on an oxygen free copper cold finger enclosed by a radiation shield, in a closed-cycle He cryostat (Montana Cryostation). The whole microscope and the cryostat are mounted on an actively damped non-magnetic optical table. Details on different possible MOI configurations can be found in Refs \onlinecite{Koblischka1995,Goa2001,Wijngaarden2001,Mandal2012}.

\section{Preparation of Co bars}
\label{sec_bars_prep}
The arrays of Co bars are defined by electron beam lithography on a Si/SiO$_2$ (300 nm) substrate covered by a resist mask (double layered, MMA/PMMA, 200 nm + 100 nm), using a nanofabrication system from Raith GmbH. After development in methyl isobutyl ketone : isopropanol (1:1) followed by an isopropanol rinse, a 30 nm thick Co layer is evaporated by molecular beam epitaxy (operation pressure $P \leq 4 \times 10^{-9}$ mbar) with a rate of $0.04$ \AA /s. A 20 nm thick Au layer is subsequently evaporated at a rate of $0.23$ \AA /s. After lift-off in acetone using a sonic bath, a 5 nm thick layer of Ti and a 45 nm thick Au layer are deposited (at rates of 0.4 \AA /s and 0.23 \AA /s respectively) on top of the structure to prevent further oxidation of the Co over the course of the experiments. All the arrays used in this study are fabricated on the same substrate. This ensures their observation under similar experimental conditions.

\end{document}